 \documentclass[prc,superscriptaddress,unsortedaddress,twocolumn,showpacs,preprintnumbers,amsmath,amssymb,floatfix,dvipdfmx]{revtex4}
\usepackage{amsmath}
\usepackage{amssymb}
\usepackage{mathrsfs}
\usepackage{multirow}
\usepackage{bm}
\usepackage{ulem}
\usepackage{color}
\usepackage[dvipdfmx]{graphicx}
\def\la{\mathrel{\mathpalette\fun <}}
\def\ga{\mathrel{\mathpalette\fun >}}
\def\fun#1#2{\lower3.6pt\vbox{\baselineskip0pt\lineskip.9pt
\ialign{$\mathsurround=0pt#1\hfil##\hfil$\crcr#2\crcr\sim\crcr}}}

\newcommand{\beq}{\begin{equation}}
\newcommand{\eeq}{\end{equation}}
\newcommand{\bea}{\begin{eqnarray}}
\newcommand{\eea}{\end{eqnarray}}

\newcommand{\bfi}[1]{\mbox{\boldmath $#1$}}
\newcommand{\bfis}[1]{\mbox{\boldmath ${\scriptstyle #1}$}}

\newcommand{\vk}{{\bfi k}}

\newcommand{\vs}{{\bfi s}}

\newcommand{\vrr}{{\bfi r}}
\newcommand{\vR}{{\bfi R}}

\newcommand{\vik}{{\bfis k}}

\def\bk{\mbox{\boldmath $k$}}

\usepackage{dcolumn}
\usepackage{lscape}

\newcolumntype{.}{D{.}{.}{-1}}
\begin{document}

\title{
Effects of chiral three-nucleon forces 
on $^{4}$He-nucleus scattering \\ 
in a wide range of incident energies}

\author{Masakazu Toyokawa}
\email[]{toyokawa@phys.kyushu-u.ac.jp}
\affiliation{Department of Physics, Kyushu University, Fukuoka 819-0395, Japan}

\author{Masanobu Yahiro}
%\email[]{yahiro@phys.kyushu-u.ac.jp}
\affiliation{Department of Physics, Kyushu University, Fukuoka 819-0395, Japan}

\author{Takuma Matsumoto}
%\email[]{matsumoto@phys.kyushu-u.ac.jp}
\affiliation{Department of Physics, Kyushu University, Fukuoka 819-0395, Japan}

\author{Michio Kohno}
%\email[]{kohno@kyu-dent.ac.jp}
\affiliation{Research Center for Nuclear Physics (RCNP), Osaka
University, Ibaraki 567-0047, Japan}

\date{\today}

\begin{abstract} 
\noindent 
{\bf Background:}
It is a current important subject 
to clarify properties of chiral three-nucleon forces 
(3NFs) not only in nuclear matter but also 
in scattering between finite-size nuclei. 
Particularly for the elastic scattering, this study has just started 
and the properties are not understood in a wide range of incident energies 
($E_{\rm in}$). \\
{\bf Aims and approach:}
We investigate basic properties of chiral 3NFs 
in nuclear matter with positive energies 
by using the Brueckner-Hartree-Fock method 
with chiral two-nucleon forces at N$^{3}$LO and 3NFs at NNLO, 
and analyze effects of chiral 3NFs 
on $^{4}$He elastic scattering 
from targets $^{208}$Pb, $^{58}$Ni and $^{40}$Ca over 
a wide range of $30 \la E_{\rm in}/A_{\rm P} \la 200$~MeV 
by using the $g$-matrix folding model, 
where $A_{\rm P}$ is the mass number of the projectile. \\
{\bf Results:} 
In symmetric nuclear matter with positive energies, 
chiral 3NFs make the single-particle potential less attractive and 
more absorptive. 
The effects mainly come from the 
Fujita-Miyazawa 2$\pi$-exchange 3NF and slightly become larger 
as $E_{\rm in}$ increases.
These effects persist in the optical potentials of 
$^{4}$He scattering. 
As for the differential cross sections of $^{4}$He scattering, 
chiral-3NF effects are large in  $E_{\rm in}/A_{\rm P} \ga 60$ MeV 
and improve the agreement of the theoretical results with the measured ones. 
Particularly in $E_{\rm in}/A_{\rm P} \ga 100$ MeV, 
the folding model  reproduces measured differential cross sections 
pretty well. 
Cutoff ($\Lambda$) dependence  is investigated 
for both nuclear matter and $^{4}$He scattering by considering 
two cases of $\Lambda=450$ and $550$~MeV. 
The uncertainty coming from the dependence is smaller than chiral-3NF effects 
even at $E_{\rm in}/A_{\rm P}=175$ MeV. 
\\ 
\end{abstract}

%21.30.Fe  Forces in hadronic systems and effective interactions
%21.65.-f  Nuclear matter
%24.10.Ht  Optical and diffraction models
%25.40.Cm  Elastic proton scattering
%25.55.Ci  Elastic and inelastic scattering(3H-,3He-, and 4He-induced reactions)
%\subjectindex{D06, D21, D22}
%D06: Effective interactions in nuclear system
%D21: Models of nucelar reactions
%D21: Light ion reactions

\maketitle

%Introduction
\section{Introduction}
\label{Introduction}
How do three-nucleon forces (3NFs) work in nuclear many-body systems? 
This is an important subject to be answered in nuclear physics. 
Even if 3NFs do not exist on a fundamental level, 
they come out in effective theories with a finite momentum cutoff 
$\Lambda$ by renormalizing the degrees of freedom 
present above $\Lambda$. 
The representative example is the 2$\pi$-exchange process 
with intermediate nucleon excited states, typically the $\Delta(1232)$ isobar. 
It is now called the Fujita-Miyazawa 3NF \cite{Fuj57}. 
As a phenomenological approach, attractive 3NFs were introduced to reproduce the binding energies 
for light nuclei \cite{Wiringa2002}, 
whereas repulsive 3NFs were used to explain the empirical saturation properties in symmetric
nuclear matter \cite{Wiringa1988}. 

%----------------------
% Figure Ch-3NF diagram
%----------------------
\begin{figure}[tbp]
\begin{center}
 \includegraphics[width=0.48\textwidth,clip]{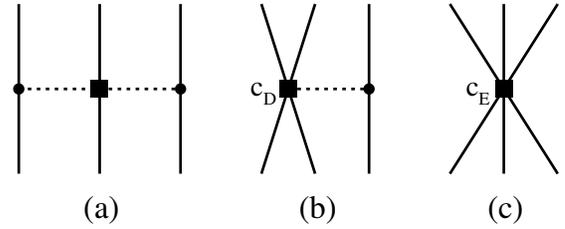}
 \caption{3NFs in NNLO. 
Diagram (a) corresponds 
to the Fujita-Miyazawa 2$\pi$-exchange 3NF \cite{Fuj57}, 
and diagrams (b) and (c) correspond to 1$\pi$-exchange and contact 3NFs.
The solid and dashed lines denote nucleon and pion propagations, 
respectively, and filled circles and squares stand for vertices. 
The strength of the filled-square vertex is often called $c_{D}$ 
in diagram (b) and $c_{E}$ in diagram (c). 
}
 \label{fig:diagram}
\end{center}
\end{figure}
%----------------------

Essential progress on this subject was made 
by chiral effective field theory (EFT) 
\cite{Epelbaum-review-2009,Machleidt-2011} 
based on chiral perturbation theory. 
The theory provides a low-momentum expansion of two-nucleon force (2NF), 
3NF and many-nucleon forces, and 
makes it possible to define the forces systematically. 
Figure \ref{fig:diagram} shows chiral 3NFs in the next-to-next-to-leading 
order (NNLO). 
Diagram (a) corresponds 
to the Fujita-Miyazawa 2$\pi$-exchange 3NF \cite{Fuj57}, and 
diagrams (b) and (c) mean 1$\pi$-exchange and contact 3NFs, respectively. 
The filled-square vertex has a strength $c_D$ 
in the diagram (b) and $c_E$ in the diagram (c). 
Quantitative roles of chiral 3NFs were extensively investigated, particularly for 
light nuclei and nuclear matter \cite{Hammer13}; more precisely, see 
Ref.~\cite{Kalantar-2012} for light nuclei, 
Refs.~\cite{Holt14,Ekstrom:2015rta} for {\it ab initio} nuclear-structure calculations in lighter nuclei and Refs.~\cite{HEB11,Samm12,Koh13,Koh15,Kru13,Dri14,Kru15} 
for nuclear matter. In addition, effects of chiral four-nucleon forces 
were found to be small in nuclear matter \cite{Kai12,Kaiser:2015lsa}. 
The chiral $g$ matrix, 
calculated from chiral 2NF+3NF with the Brueckner-Hartree-Fock (BHF) method, 
yields a reasonable nuclear matter saturation curve for 
symmetric nuclear matter, 
when the parameters, $c_{D}$ and $c_{E}$, of NNLO 3NFs are 
tuned \cite{Koh15}.

Nuclear scattering is  another place to investigate 3NF effects. The theoretical 
description of $N$+$d$ scattering has been naturally associated with the necessity of 
3NFs \cite{Kalantar-2012,Sekig14}, when the theory starts with sophisticated 2NFs determined from 
the experiments. 
Microscopic evaluation of nuclear optical potentials for 
nucleon-nucleus (NA) and nucleus-nucleus (AA) elastic scattering has 
a long history. 
The $g$-matrix folding model 
\cite{Brieva-Rook,Amos,CEG07,Raf13,MP,Toyokawa:2015zxa} 
is a standard method for deriving the optical 
potentials of NA and AA elastic scattering microscopically. 
In fact, the potentials have been used 
to analyze various kinds of nuclear reactions in many papers. 
In the model, the optical potentials were obtained 
by folding the $g$ matrix 
\cite{Brieva-Rook,Amos,CEG07,Raf13,MP,Toyokawa:2015zxa} 
with the projectile (P) density $\rho_{\rm P}$ 
and the target (T) one $\rho_{\rm T}$. 
This description has been quite successful in explaining many elastic scattering. 
At first, the effects of 3NFs were phenomenologically investigated in Ref. \cite{Raf13} 
for NA elastic scattering and in Refs. \cite{CEG07,CEG07-2} for NA and AA elastic scattering. 
The 3NFs reduce differential cross section and 
improve the agreement with measured vector analyzing powers. 
However, the role of 3NFs has not been clarified quantitatively, because 
the folding potential is adjusted to measured cross sections.

In Refs. \cite{Toyokawa:2014yma,Minomo:2014eqa}, 
as the first attempt, we made 
qualitative discussion for chiral-3NF effects on 
elastic scattering by using the hybrid method 
in which the existing local version of Melbourne $g$ matrix~\cite{Amos} 
was modified 
on the basis of the chiral $g$ matrix constructed from chiral 2NFs and 3NFs. 
The work showed that chiral-3NF effects are small 
for NA elastic scattering, but important for AA elastic scattering. 
Recently, we directly parameterized the chiral $g$ matrix 
as a local potential based on chiral 2NF+3NF, 
as briefly reported in Ref.~\cite{Toyokawa:2015zxa}. 
In this paper, we present a full understanding of chiral-3NF effects on 
${^4}$He elastic scattering over 
a wide range of $30 \la E_{\rm in}/A_{\rm P} \la 200$ MeV
by using the local version of the chiral $g$ matrix
, where $E_{\rm in}$ stands for an incident energy 
in the laboratory system and $A_{\rm P}$ is the mass number of projectile.

The $g$ matrices calculated so far are provided by a local potential with 
Yukawa or Gaussian form, since this procedure makes the folding 
calculation much easier.

Investigation of chiral-3NF effects on NA and AA elastic scattering 
has just started with lower incident energies per nucleon such as 
$E_{\rm in}/A_{\rm P} \approx 70$ MeV 
by using the $g$-matrix folding model 
\cite{Minomo:2014eqa,Toyokawa:2014yma,Toyokawa:2015zxa}, 
since chiral EFT is more reliable for lower incident energies. 
As mentioned above, the folding potentials were recently calculated 
from the local version  of chiral $g$ matrix in Ref. \cite{Toyokawa:2015zxa}. 
The chiral $g$-matrix folding model accounts for experimental data 
considerably well on NA scattering at $E_{\rm in}=65$ MeV and $^{4}$He+$^{58}$Ni scattering at 
$E_{\rm in}/A_{\rm P}=72$ MeV. 
This model also showed that chiral-3NF effects are small for NA elastic scattering, 
but sizable for $^{4}$He elastic scattering.

In our previous studies for $^4$He elastic scattering, we used the Melbourne g matrix in 
Ref. \cite{Egashira:2014zda} and the chiral g matrices based on chiral 2NF and chiral 2NF+3NF 
in Ref. \cite{Toyokawa:2015zxa}. 
After Ref. \cite{Toyokawa:2015zxa} was published, we found some numerical errors in our 
nuclear-matter calculations including chiral 3NFs; see Ref. \cite{Koh:Err} for the details. 
In the present work, we then adopt the corrected version of chiral $g$-matrix; see Appendix for the matrix. 
Further discussion will be made later in Sec. \ref{Some basic results of BHF calculations}.

In this paper, 
we first investigate basic properties of chiral 3NFs in symmetric 
nuclear matter {\it for positive energies} up to 200 MeV 
by using the BHF method with chiral 2NFs of N$^{3}$LO 
and chiral 3NFs of NNLO. 
We show that chiral-3NF effects provide density-dependent repulsive and absorptive corrections 
to the single-particle potential and that the effects slightly become larger as 
the energy increases. 
We also point out that the corrections mainly come 
from the Fujita-Miyazawa 2$\pi$-exchange 3NF of diagram (a).

Second, we analyze  chiral-3NF effects on $^{4}$He  
scattering from various targets 
in a {\it wide range of incident energies} 
by using the chiral $g$-matrix folding model. 
In order to make our discussion clear, 
we take $^{4}$He scattering as AA scattering, 
since the $g$-matrix folding model is confirmed 
to work well for $^{4}$He scattering in virtue of 
negligibly small 
projectile-breakup effects~\cite{Egashira:2014zda,Toyokawa:2015fva}; 
see Sec. \ref{Folding model} for further discussion. 
In addition, as targets we take heavier nuclei, 
$^{208}$Pb, $^{58}$Ni and $^{40}$Ca, 
since the $g$ matrix is evaluated in nuclear matter 
and is considered to be more 
suitable for heavier targets. 
For the targets, 
the experimental data are available in a wide range of 
$30 \la E_{\rm in}/A_{\rm P} \la 200$~MeV.

In the present paper, we mostly consider the case of the cutoff scale 
$\Lambda=550$ MeV. As the third subject, $\Lambda$ dependence 
is investigated for nuclear matter with positive energies and 
$^{4}$He elastic scattering 
by taking two other cases of $\Lambda=450$ and $550$ MeV.

Finally, we provide the local version of chiral $g$ matrix including 
chiral-3NF effects with a 3-range Gaussian form for the case of 
$E_{\rm in}/A_{\rm P}=75$ MeV. This may strongly encourage 
the application of the chiral $g$ matrix for studying various kinds 
of nuclear reactions. 
This local version of chiral $g$ matrix is referred to as 
\lq\lq Kyushu chiral $g$ matrix'' in this paper.

In Sec. \ref{Theoretical framework}, we 
present  the theoretical framework composed of the BHF method 
and the folding model, and show some basic results of BHF calculations 
for chiral 2NF+3NF. 
In Sec. \ref{Results}, the results of 
the chiral g-matrix folding model are shown 
for $^{4}$He elastic scattering. 
Section \ref{Summary} is devoted to a summary.

%Theoretical framework
\section{Theoretical framework and basic results}
\label{Theoretical framework}

\subsection{BHF equation for 2NF+3NF} 
\label{BHF equation for 2NF+3NF}

We first recapitulate the BHF method for 2NF+3NF, 
following Ref. \cite{Koh13}. 
Because it is not easy to treat a 3NF $V_{123}$ even in nuclear matter, 
we introduce an effective 2NF $V_{12}^{\rm eff}$ 
by applying the mean-field approximation, or the normal ordering prescription, 
to the 3NF: 
\begin{eqnarray}
 &&
 \frac{1}{2} \sum_{\vik_1 \vik_2} \langle \bk_1 \bk_2 |
  V_{12} |\bk_1 \bk_2\rangle_{\cal A} 
  \nonumber\\
 &&~~~
 +\frac{1}{3!}\sum_{\vik_1 \vik_2 \vik_3} \langle \bk_1 \bk_2 \bk_3| 
  V_{123} |\bk_1 \bk_2 \bk_3\rangle_{\cal A}
  \nonumber\\
 &&=\frac{1}{2} \sum_{\vik_1 \vik_2} \langle \bk_1 \bk_2 | 
  V_{12}^{\rm eff}
|\bk_1 \bk_2\rangle_{\cal A}  , 
\label{pot-energy}
\end{eqnarray} 
where ${\cal A}$ means the antisymmetrization and 
$\bk_i$ corresponds to quantum numbers of the $i$-th nucleon. 
Equation \eqref{pot-energy} leads
\bea
 V_{12}^{\rm eff}=V_{12}+ \frac{1}{3}V_{12(3)} , 
\label{eff-V12}
\eea
where  $V_{12(3)}$ is defined by summing up 
3NF $V_{123}$ over the third nucleon in the Fermi sea:
\begin{equation}
  \langle \bk_1' \bk_2' |  V_{12(3)} |\bk_1 \bk_2\rangle_{\cal A}
   = \sum_{\bk_3} \langle \bk_1' \bk_2' \bk_3 |
  V_{123} |\bk_1 \bk_2 \bk_3 \rangle_{\cal A} 
\end{equation}
with assuming the center-of-mass (c.m.) frame: $\bk_1'+\bk_2'=\bk_1+\bk_2$. 
Note {\it the factor $1/3$ } in Eq. \eqref{eff-V12}. 
The $g$ matrix $g_{12}$ is a solution to the BHF equation 
\begin{equation}
 g_{12}=V_{12}^{\rm eff}+V_{12}^{\rm eff} G_0 g_{12} , 
\label{g-eq}
\end{equation}
where $G_0$ is the nucleon propagator with the Pauli exclusion operator 
in the numerator and with the single-particle energy 
\bea
e_{\vik}=\langle \bk |T|\bk\rangle + {\rm Re}[{\cal U}(\bk)]  
\eea
of the nucleon having a momentum $\vk$ 
in the denominator. Here $T$ is the standard kinetic-energy operator 
of nucleon, and the single-particle potential ${\cal U}(\bk)$ is  
defined by~\cite{Koh13} 
\bea
{\cal U}(\bk) = \sum_{|\vik'|\le k_{\rm F}}
\langle \bk \bk' | \tilde{g}_{12} 
|\bk \bk' \rangle_{\cal A} . 
\label{single-particle-pot-0}
\eea
with the effective $g$ matrix, so-called $\tilde{g}$ matrix, including additional 
rearrangement terms of the 3NF origin: 
\bea
 \tilde{g}_{12}=g_{12} 
+\frac{1}{6} V_{12(3)}(1+G_0 g_{12}) .
\label{effective-g}
\eea
Note that $\vk$ is related to 
the incident energy $E_{\rm in }$ as 
$E_{\rm in }=(\hbar \vk)^2/(2m) + {\rm Re}[{\cal U}]$. 
The present formulation is consistent 
with the second-order perturbation of Ref. \cite{Hol13}, 
because of {\it  the factor $1/6$ } 
in Eq. \eqref{effective-g}. 
For the symmetric nuclear matter where 
the proton density $\rho_{p}$ agrees with the neutron one $\rho_{n}$, 
the Fermi momentum $k_{\rm F}$ is 
related to the matter density $\rho=\rho_{p}+\rho_{n}$ 
as $k_{\rm F}^3= 3\pi^2 \rho/2$, so that 
the normal density $\rho=\rho_0=0.17$ fm$^{-3}$ is realized 
at $k_{\rm F}=1.35$ fm$^{-1}$.

\subsection{Some basic results of BHF calculations} 
\label{Some basic results of BHF calculations}

The $\tilde{g}$ matrix is calculated 
from chiral 2NF of N$^{3}$LO and chiral 3NF of NNLO by using the BHF method. 
In BHF calculations, the form factor $\exp\{-(q'/\Lambda)^6-(q/\Lambda)^6\}$ is introduced for 
both $V_{12}$ and $V_{12(3)}$. 
We mainly consider the case of $\Lambda=550$ MeV, and take another case $\Lambda= 450$ MeV  
when $\Lambda$ dependence of physical quantities is estimated. 
The low-energy constants relevant for 3NFs are 
$(c_1,c_3,c_4)=(-0.81,-3.4,3.4)$~\cite{Epe05} in units of GeV$^{-1}$.

As noted earlier, some errors were found in nuclear-matter calculations with chiral 3NFs of Ref. \cite{Koh13},
after Ref. \cite{Toyokawa:2015zxa} was published. Although the qualitative importance of chiral 3NFs
for improving nuclear matter saturation properties does not change, the saturation curve is changed
by the corrections. To restore reasonable nuclear saturation properties, which are basically important
for further application for microscopic derivation of nuclear optical potentials, the remaining two parameters
$c_D$ and $c_E$ are tuned \cite{Koh:Err}. In consideration of the uncertainty that the $c_D$ and $c_E$ terms
yield almost identical contributions when $c_D \simeq 4c_E$, $c_D$ is determined as $-2.5$ by setting
$c_E=0$ for $\Lambda=450$~MeV and next $c_E$ is fixed as $0.25$ for $\Lambda=550$~MeV with keeping
$c_D=-2.5$. These values are somewhat different from those determined in few-body systems within
continuous uncertainties. It has been recognized \cite{Ekstrom:2015rta}, however, that low-energy-constants
fixed solely in few-body systems are not adequate in heavier systems. In this article, we use the corrected
version of the chiral $g$ matrix.

\begin{figure}[htbp]
\begin{centering}
 \includegraphics[width=0.38\textwidth,clip]{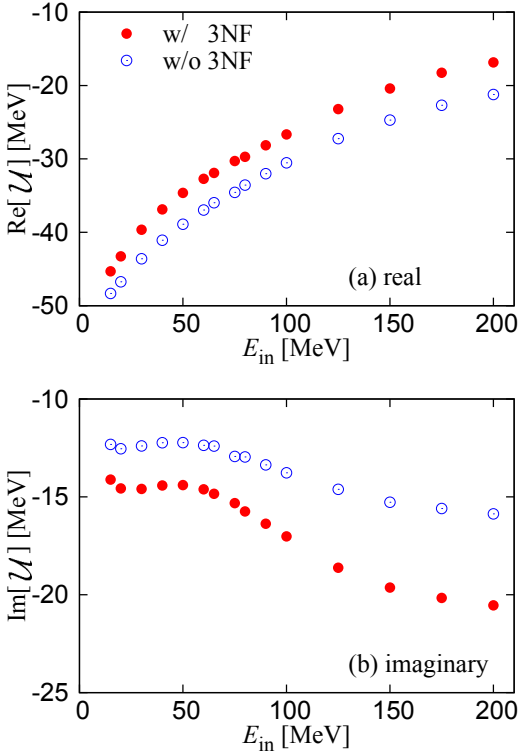}
 \caption{(Color online) 
$E_{\rm in}$ dependence of ${\cal U}$ 
at $k_{\rm F}=1.2$ fm$^{-1}$ for the cutoff $\Lambda = 550$ MeV. 
Filled (open) circles stand for the results of BHF calculations with 
(without) chiral 3NFs. Panels (a) and (b) correspond to the real and imaginary 
parts of ${\cal U}$. 
}
 \label{fig-SPP-E-dep-3NF-effect}
\end{centering}
\end{figure}

It is known that chiral 3NFs make repulsive corrections to 
the binding energy of symmetric nuclear matter~\cite{Koh13}. 
What happens in positive energy? 
Figure \ref{fig-SPP-E-dep-3NF-effect} shows $E_{\rm in}$ dependence 
of ${\cal U}$ for the case of $k_{\rm F}=1.2$ fm$^{-1}$ for the cutoff $\Lambda = 550$ MeV. 
This density is realized in the peripheral region of a target nucleus 
and hence important for elastic scattering. 
Filled  (open) circles denote the results of BHF calculations with 
(without) chiral 3NFs. 
One can see that chiral 3NFs make ${\cal U}$ less attractive and more 
absorptive. 
The 3NF corrections slightly increase as $E_{\rm in}$ goes up.
Our results are consistent with the second-order perturbation calculation by Holt 
{\it et. al.} \cite{Hol13}.

%----------------------
% Figure Ein-dep of U: E-dep of 3NF-effect
%----------------------
%----------------------

\begin{figure}[htbp]
\begin{centering}
 \includegraphics[width=0.38\textwidth,clip]{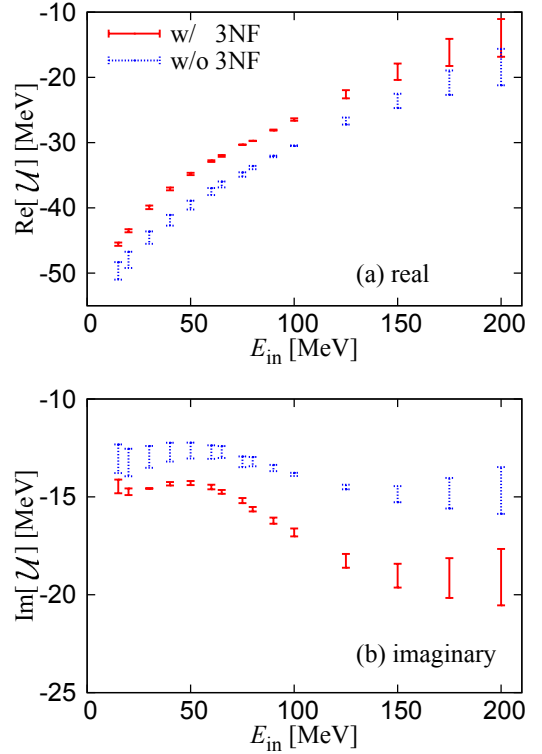}
 \caption{(Color online) 
Single-particle potential ${\cal U}$ as a function of $E_{\rm in}$  
at $k_{\rm F}=1.2$ fm$^{-1}$ for two cases of $\Lambda=450, 550$~MeV. 
$\Lambda$ dependence is shown as an error bar. 
The error bar plotted by a solid (dashed) line means 
the results of BHF calculations with (without) chiral 3NFs. 
Panels (a) and (b) mean the real and imaginary 
parts of ${\cal U}$, respectively. 
}
 \label{fig-SPP-E-dep-3NF-Lambda}
\end{centering}
\end{figure}

Figure \ref{fig-SPP-E-dep-3NF-Lambda} shows ${\cal U}$ 
as a function of $E_{\rm in}$ at $k_{\rm F}=1.2$ fm$^{-1}$, 
but two cases of $\Lambda=450$ and $550$~MeV are taken 
in BHF calculations to see the uncertainty coming from 
$\Lambda$ dependence on ${\cal U}$. 
The $\Lambda$ dependence is plotted as an error bar. 
The error bar plotted by a solid (dashed) line denotes 
the results of BHF calculations with (without) chiral 3NFs; 
note that panels (a) and (b) correspond to 
the real and imaginary parts of ${\cal U}$. 
Particularly for BHF calculations with chiral 3NFs, 
there is a tendency that the uncertainty 
become larger as $E_{\rm in}$ increases from 80~MeV. 
Even at $E_{\rm in} = 175$~MeV, however, 
chiral 3NF effects are larger than the uncertainty. 
This enables us to make reliable discussion on chiral-3NF effects.

%----------------------
% Figure : Lambda-dep and 3NF-effect
%----------------------
%----------------------

In order to obtain deeper understanding of the properties
of chiral 3NFs, 
we classify $\tilde{g}(k_{\rm F},E_{\rm in})$ with 
the total spin $S$ and isospin $T$ 
of the interacting two-nucleon system. 
The total single-particle potential ${\cal U}$ 
is obtained by 
the single-particle potential ${\cal U}^{ST}$ in each $(S,T)$ channel as 
\bea
{\cal U}=\sum_{ST}(2S+1)(2T+1){\cal U}^{ST} , 
\eea
where ${\cal U}^{ST}$ is defined by 
Eq. \eqref{single-particle-pot-0} with $\tilde{g}$ replaced by 
$\tilde{g}^{ST}$.

\begin{figure}[tbp]
\begin{centering}
 \includegraphics[width=0.40\textwidth,clip]{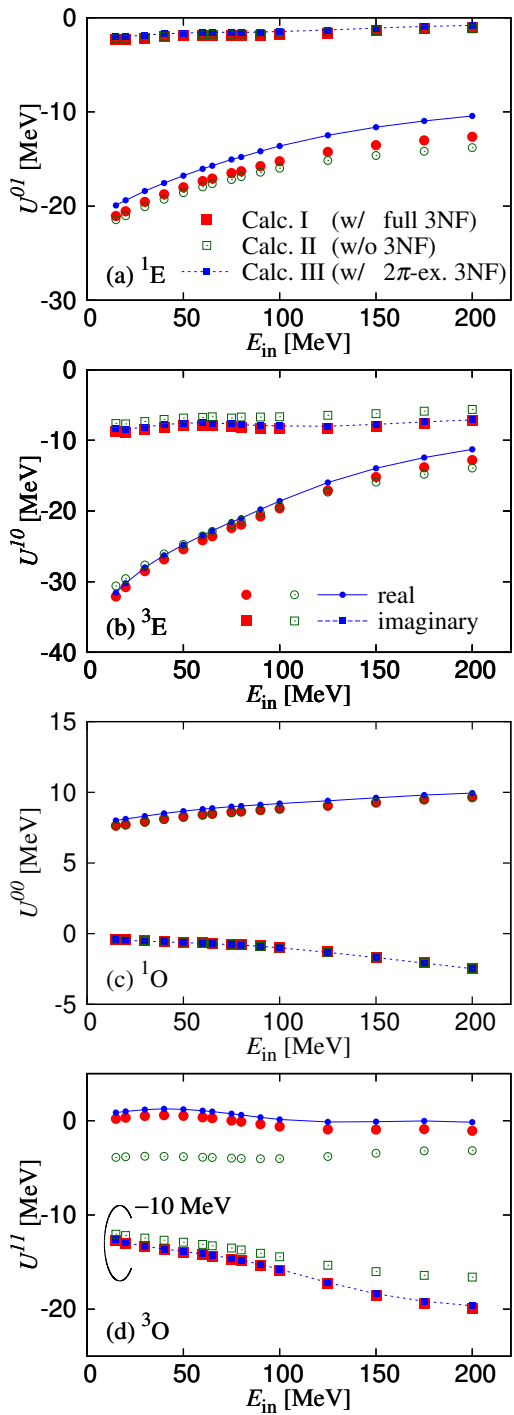}
 \caption{(Color online) 
$E_{\rm in}$ dependence of $U^{ST} \equiv (2S+1)(2T+1){\cal U}^{ST}$ 
at $k_{\rm F}=1.2$ fm$^{-1}$  
for (a)$^{1}$E ($S=0,T=1$) , (b) $^{3}$E ($S=1,T=0$), (c) $^{1}$O 
($S=0,T=0$), and (d) $^{3}$O ($S=1,T=1$). 
Filled circles (squares) represent $U^{ST}$ in its real 
(imaginary) part obtained by BHF calculations with all kinds of chiral 3NFs. 
Open circles (squares) correspond to 
the real (imaginary) part of $U^{ST}$ in which all kinds of chiral 
3NFs are switched off. 
Lines with small circles (squares) stand for $U^{ST}$ 
obtained by BHF calculations with $c_D=c_E=0$. 
Note that lines are a guide to the eye; 
the solid (dashed) line corresponds to the real (imaginary part). 
For $^{3}$O, the imaginary part is shifted down by 10 MeV. 
}
 \label{fig-SPP-ST-E-dep-FM-dominance}
\end{centering}
\end{figure}

Figure \ref{fig-SPP-ST-E-dep-FM-dominance} shows 
$E_{\rm in}$ dependence of $U^{ST}\equiv (2S+1)(2T+1){\cal U}^{ST}$ for the case of 
$k_{\rm F}=1.2$ fm$^{-1}$. 
Here we do the following three kinds of BHF calculations:

\begin{itemize}
 \item[I.] All kinds of chiral 3NFs, i.e., diagrams (a)-(c) 
in Fig. \ref{fig:diagram}, are taken into account. 
 \item[II.] All kinds of chiral 3NFs are switched off. Namely, 
Only chiral 2NF is considered. 
 \item[III.] Diagrams (b) and (c) are ignored by setting $c_D=c_E=0$ 
in BHF calculations. Namely, only the Fujita-Miyazawa 2$\pi$-exchange 3NF of 
diagram (a) is considered. 

\end{itemize}
Filled  circles (squares) stand for the real (imaginary) part of 
$U^{ST}$ for calculation I, 
while open circles (squares) correspond to 
the real (imaginary) part of $U^{ST}$ for calculation II; 
note that lines are a guide to the eye. 
The two calculations show that chiral 3NF effects are significant 
for $^{3}$O ($S=1,T=1$) and $^{3}$E ($S=1,T=0$) channel and 
the real part of $^{1}$E ($S=0,T=1$) channel. 
Small circles (squares) represent the real (imaginary) part of 
$U^{ST}$ for calculation III. 
For $^{3}$E and $^{3}$O, 
one can see from calculations II and III that chiral 3NF effects 
mainly come from the Fujita-Miyazawa 2$\pi$-exchange 3NF of diagram (a). 
For the real part of $^{1}$E ($S=0,T=1$) channel, the effect of diagram (a) 
is sizable, but it is considerably 
reduced by the effects of diagram (b) and (c).  
As a net effect of these properties, 
chiral 3NFs make ${\cal U}$ less attractive and more absorptive, 
and the repulsion mainly stems from diagram (a) in its $^{3}$O component 
and the absorption does from diagram (a) in its $^{3}$O and $^{3}$E 
components. 
The chiral-3NF effects become more significant at larger incident energies.  
One can easily expect that these properties persist 
also in the optical potentials of $^{4}$He scattering, since 
${\cal U}$ plays a role of \lq\lq optical potential'' of 
nucleon scattering in nuclear matter. 
This point will be discussed later in Sec. \ref{Results}.

%----------------------
% Figure Ein-dep of U_{ST}: FM 3NF dominance
%----------------------
%----------------------

\subsection{Local version of chiral $g$ matrix} 
\label{Local version of chiral $g$ matrix}

The $\tilde{g}$ matrix  $\tilde{g}(k_{\rm F},E_{\rm in})$ of
Eq.\eqref{effective-g} is a nonlocal potential 
depending on $k_{\rm F}$ and $E_{\rm in}$, being calculated in symmetric nuclear matter. 
In addition, it is obtained numerically. 
These properties are quite inconvenient in various applications. 
In order to circumvent the problem, the Melbourne group showed 
that elastic scattering are determined by 
the on-shell and near-on-shell components of $g$ matrix \cite{Amos}, 
and provided a local version of $g$ matrix in which 
the potential parameters 
are so determined as to reproduce the relevant components 
\cite{Amos,von-Geramb-1991,Amos-1994}. 
The Melbourne $g$ matrix thus obtained well accounts for NN scattering 
in free space that corresponds to the limit of $\rho=0$, 
and the Melbourne $g$-matrix folding model reproduces NA scattering, 
as already mentioned in Sec. \ref{Introduction}.

In our previous paper \cite{Toyokawa:2015zxa}, 
following the Melbourne-group procedure \cite{von-Geramb-1991,Amos-1994,Amos},
we succeeded in parameterizing
a local version of chiral $\tilde{g}$ matrix in a 3-range Gaussian form 
for each of the central, spin-orbit and tensor components. 
The Gaussian form makes various kinds of numerical calculations efficient. 
The range and strength parameters were so determined 
as to reproduce the on-shell and near-on-shell matrix elements 
of the original $\tilde{g}$ matrix for each spin-isospin channel, 
$k_{\rm F}$ and 
$E_{\rm in}$. As for the central part, 
the range parameters obtained were $(0.4,0.9,2.5)$ in units of fm. 
In this paper, we repeated this procedure for $E_{\rm in}$ up to 200 MeV and 
parameterized a local version of chiral $\tilde{g}$ matrix with good accuracy, 
as shown below. 
Since the analysis was already made at $E_{\rm in}=65$ MeV 
in Ref. \cite{Toyokawa:2015zxa}, we make the same analysis 
for higher energies, say $E_{\rm in}=150$ MeV, in this paper. 
Whenever we have to distinguish the two types of $g$ matrices, 
we call the local version of 
$\tilde{g}$ matrix \lq\lq Kyushu chiral $g$ matrix'' 
and the original nonlocal $\tilde{g}$ matrix  
\lq\lq original chiral $g$ matrix''. 
For the case of $E_{\rm in}=75$ MeV as an example, we present the parameter set
of Kyushu chiral $g$ matrix in Appendix \ref{Appendix}.

%----------------------
% Figure p+n
%----------------------
\begin{figure}[tbp]
\begin{centering}
 \includegraphics[width=0.40\textwidth,clip]{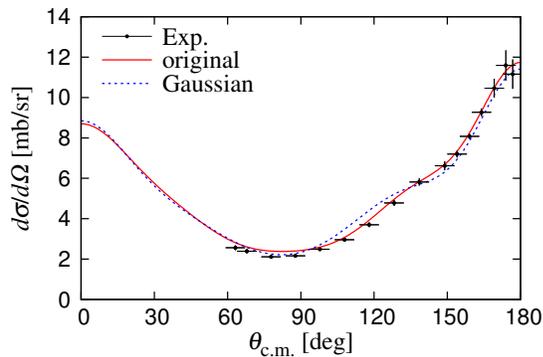}
 \caption{(Color online) 
Differential cross sections for $p$+$n$ scattering at 
$E_{\rm in } = 150$ MeV in free space. Here 
$\theta_{\rm c.m.}$ denotes the scattering angle in the center of mass system. 
The solid line stands for  the result of 
original chiral $t$ matrix, while 
the dashed line corresponds to the result of 
Kyushu chiral $t$ matrix (the local version of chiral $t$ matrix). 
Experimental data are taken from Ref.~\cite{np:Measday}. 
}
 \label{fig:n+p-scattering}
\end{centering}
\end{figure}
%----------------------

Figure \ref{fig:n+p-scattering} shows differential cross sections 
as a function of c.m. scattering angle 
$\theta_{\rm c.m.}$ 
for $p$+$n$ scattering at $E_{\rm in } =150$ MeV 
in free space, i.e., in the limit of $\rho=0$. 
The solid and dashed lines denote the results of 
original and Kyushu chiral $t$ matrices, respectively; 
note that the $g$ matrix is reduced to the $t$ matrix 
in the limit of $\rho=0$.  
The Kyushu chiral $t$ matrix reproduces the result 
of original chiral $t$ matrix well.

\begin{figure}[tbp]
\begin{centering}
 \includegraphics[width=0.40\textwidth]{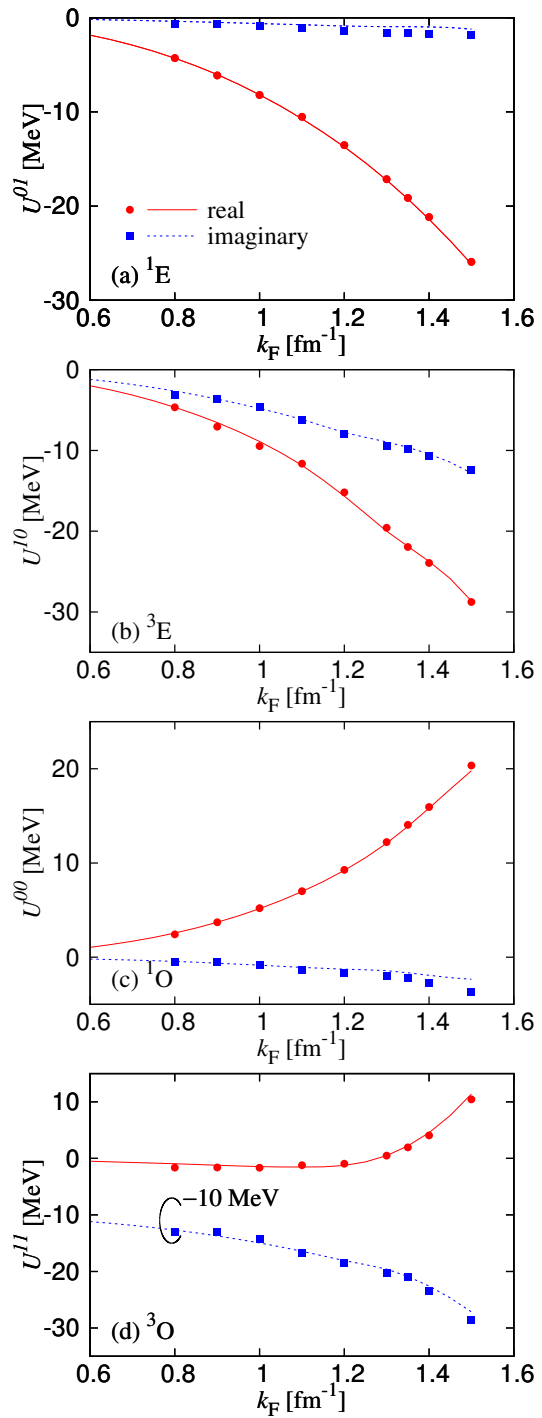}
\caption{(Color online) 
$k_{\rm F}$ dependence of $U^{ST}$ at $E_{\rm in}=150$ MeV  
for (a)$^{1}$E, (b) $^{3}$E, (c) $^{1}$O, and (d) $^{3}$O. 
Here 3NFs are taken into account in BHF calculations. 
The filled circles (squares) stand for the results of the real 
(imaginary) part of original chiral $g$ matrix, 
while the solid (dashed) lines correspond to the results of 
the real (imaginary) part of Kyushu chiral $g$ matrix. 
For $^{3}$O, the imaginary part is shifted down by 10 MeV. 
}
\label{fig-SPP-ST}
\end{centering}
\end{figure}

Figure \ref{fig-SPP-ST} shows  $k_{\rm F}$ dependence of $U^{ST}$ 
at $E_{\rm in}=150$ MeV. Both 2NF and 3NF are taken into account in BHF calculations. 
The filled circles (squares) denote the results of the real 
(imaginary) part of original chiral $g$ matrix, 
whereas 
the solid (dashed) lines correspond to 
the real (imaginary) part of Kyushu chiral $g$ matrix. 
The range $k_{\rm F} \la 1.35$ fm$^{-1}$ ($\rho \la \rho_0$) contributes to 
the optical potentials of $^{4}$He scattering, 
when the potentials are constructed by the folding model explained in 
Sec. \ref{Folding model}. 
In particular, the Fermi momentum $k_{\rm F} \approx 1.2$ fm$^{-1}$, 
corresponding to the peripheral region of the optical potentials, is 
important for the elastic scattering. 
The Kyushu chiral $g$ matrix well 
reproduces the results of the original chiral $g$ matrix.

%----------------------
% Figure single particle potential
%----------------------
%----------------------

\subsection{Folding model} 
\label{Folding model}

In this paper, the optical potentials are derived 
by folding Kyushu chiral $g$ matrix 
with $\rho_{\rm P}$ and $\rho_{\rm T}$ 
for $^{4}$He scattering on $^{208}$Pb, $^{58}$Ni and $^{40}$Ca 
targets. 
In general, the folding potential is referred to as 
a double-folding (DF) model for AA scattering, while it is called 
a single-folding (SF) model for NA scattering. 

In the $g$-matrix SF model for NA elastic scattering, 
the so-called local-density approximation is taken, 
that is, the value of $\rho$ in $g(\rho)$ is identified 
with the value of $\rho_{\rm T}$ 
at the midpoint $\vrr_{\rm m}$ of interacting two nucleons: 
$\rho=\rho_{\rm T}(\vrr_{\rm m})$. 
Target-excitation effects on the elastic scattering are 
well taken into account by this framework. 
In fact, the Melbourne $g$-matrix SF model succeeded in reproducing 
NA scattering \cite{Amos}. 
In our previous work \cite{Toyokawa:2015zxa}, furthermore, we showed that 
the Kyushu chiral $g$-matrix SF model also well accounted for proton 
scattering at $E_{\rm in}=65$~MeV 
and chiral-3NF effects are small there.

The $g$-matrix DF model for AA scattering had a problem to be settled. 
In order to obtain the $g$ matrix applicable for AA scattering, 
in principle, we have to consider two Fermi spheres in 
nuclear-matter calculations and solve a collision between a nucleon 
in the first Fermi sphere and a nucleon in the second 
one~\cite{Izu80,Yahiro-Glauber}. 
However, actual calculations are not feasible. 
In fact, all the $g$ matrices provided so far were obtained 
by assuming a single Fermi sphere and solving nucleon scattering 
on the Fermi sphere. 
For consistency with the nuclear-matter calculation, 
we assumed $\rho=\rho_{\rm T}(\vrr_{\rm m})$ in $g(\rho)$ 
and applied the framework to $^{3,4}$He scattering 
in a wide energy range of $30 \la E_{\rm in}/A_{\rm P} \la 180$ MeV 
\cite{Egashira:2014zda,Toyokawa:2015fva}. 
The Melbourne $g$-matrix DF model based on the target-density approximation (TDA) 
well accounted for $^{3,4}$He scattering, 
particularly for forward differential cross sections where 3NF effects are 
considered to be negligible~\cite{CEG07,CEG07-2,Raf13}. 
In our previous analysis \cite{Toyokawa:2015zxa}, 
the DF-TDA model based on Kyushu chiral $g$ matrix well 
explained $^{4}$He scattering at $E_{\rm in}/A_{\rm P} \approx 72$ MeV. 
We then take the DF-TDA model for $^{4}$He scattering in this paper 
throughout all the incident energies 
$30 \la E_{\rm in}/A_{\rm P} \la 180$ MeV where the experimental data 
are available.

The DF model naturally treats both the direct and knock-on exchange processes 
\cite{Yahiro-Glauber,Tang1978,Aoki1982}. 
In the latter process, interacting two nucleons are exchanged and thereby 
the potential becomes nonlocal. However, the nonlocality can be 
localized with high accuracy by the local momentum approximation 
\cite{Brieva-Rook}, as proven in Refs.~\cite{Hag06,Minomo:2009ds}. 
The folding potential $U(R)$ thus obtained 
is a function of the distance $R$ between P and T;
\begin{eqnarray}
 &&U(R)=
  \sum_{\mu\nu}\int d\vrr_{\rm P} \int d\vrr_{\rm T}~
  \rho_{\rm P}^{(\mu)}(r_{\rm P})\rho_{\rm T}^{(\nu)}(r_{\rm T})
  \nonumber\\
 &&~~~~~~~~~~~\times
  \tilde{g}^{\rm DR}_{\mu\nu}(s,E_{\rm in}/A_{\rm P};\rho)
  \nonumber\\
 &&~~~~~~-\sum_{\mu\nu}\int d\vrr_{\rm P} \int d\vrr_{\rm T}~
  \tilde{\rho}_{\rm P}^{(\mu)}(\vrr_{\rm P},\vs)  \tilde{\rho}_{\rm T}^{(\nu)}(\vrr_{\rm T},\vs)
  \nonumber\\
 &&~~~~~~~~~~~\times
  \tilde{g}^{\rm EX}_{\mu\nu}(s,E_{\rm in}/A_{\rm P};\rho)
  j_{0}({\scriptstyle \frac{A_{\rm P}+A_{\rm T}}{A_{\rm P}A_{\rm T}}}K(R)s), 
\end{eqnarray}
where the indices $\mu$ and $\nu$ are the isospin of corresponding nucleon and 
$\vs= \vrr_{\rm T}-\vrr_{\rm P}-\vR$ is the coordinate between interacting two nucleons. 
The densities $\rho_{\rm P(T)}$ and $\tilde{\rho}_{\rm P(T)}$ represent 
the one-body and mixed densities of P (T); 
\begin{eqnarray}
 \tilde{\rho}_{\rm P(T)}&=&
  \rho_{\rm P(T)}(|\vrr_{\rm P(T)}\pm \vs/2|)
  \frac{3j_{1}(k_{\rm F}^{\rm P(T)}s)}{k_{\rm F}^{\rm P(T)}s}
  . 
\end{eqnarray}
The Fermi momentum $k_{\rm F}^{\rm P(T)}$ is related to the density $\rho_{\rm P(T)}$. 
The direct (exchange) term of $g$-matrix $\tilde{g}^{\rm DR(EX)}_{\mu \nu}$ is defined by 
$\tilde{g}^{ST}$ as 
\begin{eqnarray}
 \tilde{g}^{\rm DR(EX)}_{pp,nn}
  &=&\frac{1}{4}(\pm \tilde{g}^{01}+3\tilde{g}^{11})~, 
  \\
 \tilde{g}^{\rm DR(EX)}_{pn,np}
  &=&\frac{1}{8}(\tilde{g}^{00}\pm \tilde{g}^{01} \pm 3\tilde{g}^{10} + 3\tilde{g}^{11})~. 
\end{eqnarray}
See Refs.~\cite{CEG07-2,Khoa94,Khoa97,Toyokawa:2015fva} for the detail of the formulation 
of the DF model. 
The $S$ matrices for $^{4}$He elastic scattering are obtained by solving 
the one-body Schr\"odinger equation with $U(R)$.

For the targets $^{208}$Pb and $^{58}$Ni, 
the matter densities $\rho_{\rm T}$ are evaluated 
by the spherical Hartree-Fock (HF) method 
based on the Gogny-D1S interaction~\cite{GognyD1S}, where 
the spurious c.m. motions are 
removed with the standard manner~\cite{Sum12}. 
For the projectile $^{4}$He and  the target $^{40}$Ca,  
we take the phenomenological proton-density 
determined from electron scattering~\cite{phen-density}; 
here the finite-size effect of proton charge 
is unfolded with the standard procedure~\cite{Singhal}, and 
the neutron density is assumed to have the same geometry as the 
proton one, since the difference between the neutron root-mean-square radius 
and the proton one is only 1\% in spherical HF calculations.

\section{Results}
\label{Results}

Now we  analyze {$^{4}$He elastic scattering on nuclei systematically 
in a wide range $E_{\rm in}/A_{\rm P}=26$--$175$ MeV}. 
Here heavier targets $^{208}$Pb, $^{58}$Ni and $^{40}$Ca are considered, 
because the $g$ matrix is calculated in nuclear matter and thereby 
the $g$-matrix DF model is expected to be more reliable for heavier targets.

\begin{figure*}[tbp]
\begin{centering}
 \includegraphics[width=0.86\textwidth,clip]{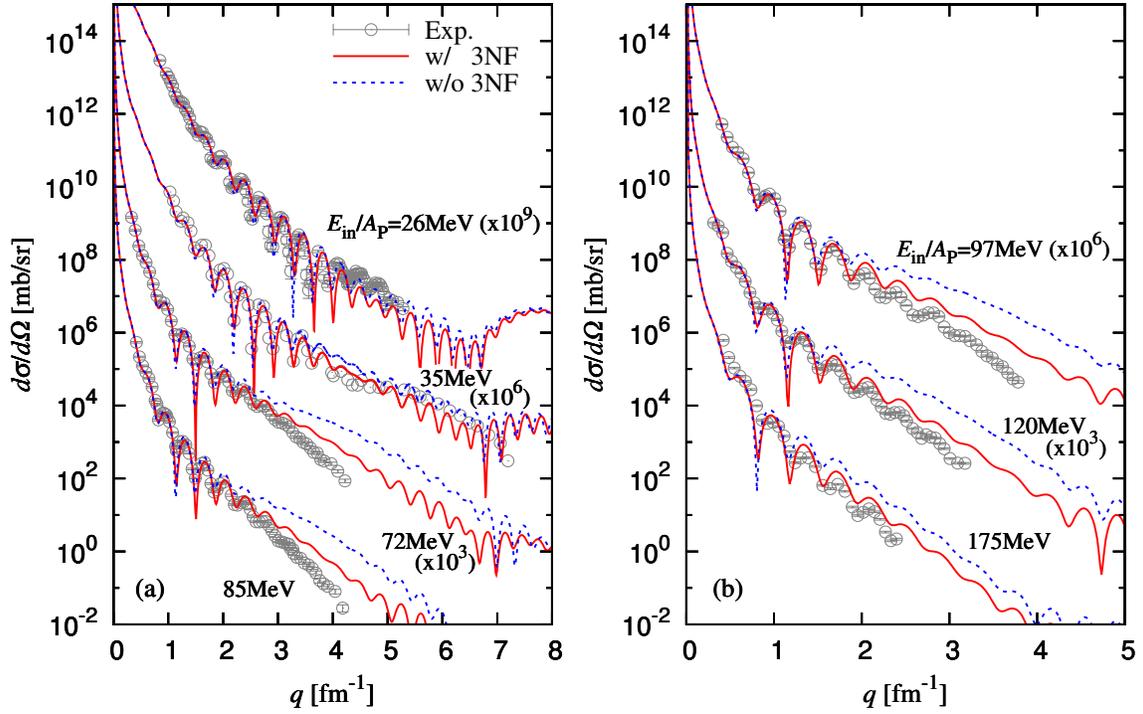}
 \caption{(Color online) 
Differential cross sections $d \sigma/d \Omega$ 
as a function of transfer momentum $q$ 
for $^{4}$He scattering from a $^{208}$Pb target at 
$E_{\rm in}/A_{\rm P}=26$--$175$ MeV.  
The solid (dashed) lines denote the results of 
Kyushu chiral $g$ matrix with (without) 3NF effects. 
Each cross section is multiplied by the factor shown in the figure.  
Experimental data are taken from Refs.~\cite{He4Ca40Pb208:Hauser,
He4Pb208:Goldberg,He4Ni58Pb208:Bonin,He4Pb208:Uchida}. 
}
 \label{fig:4He+208Pb-scattering}
\end{centering}
\end{figure*}

\begin{figure*}[tbp]
\begin{centering}
 \includegraphics[width=0.86\textwidth,clip]{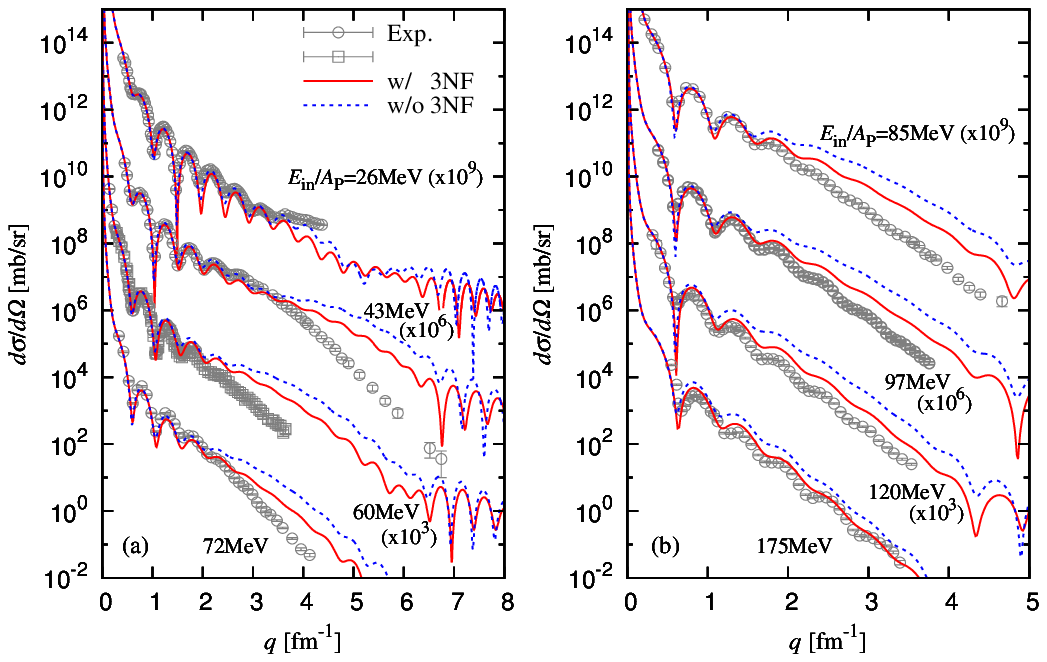}
 \caption{(Color online) 
Same as Fig. \ref{fig:4He+208Pb-scattering}, but the target nucleus 
is $^{58}$Ni. 
Experimental data are taken from Refs.~\cite{He4Ni58:Rebel,
He4Ni58:Albinski,He4Ni58:Clark,He4Ni58:Lui,He4Ni58Pb208:Bonin,
He4Ni58:Nayak}. 
}
 \label{fig:4He+58Ni-scattering}
\end{centering}
\end{figure*}

\begin{figure*}[tbp]
\begin{centering}
 \includegraphics[width=0.43\textwidth,clip]{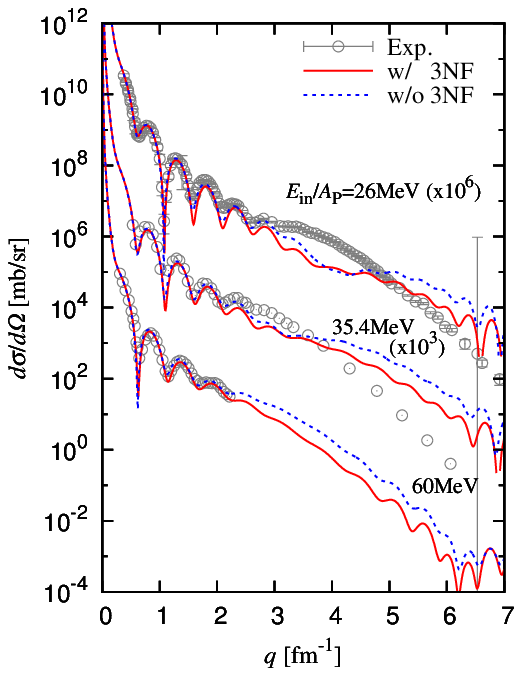}
 \caption{(Color online) 
Same as Fig. \ref{fig:4He+208Pb-scattering}, but the target nucleus 
is $^{40}$Ca. 
Experimental data are taken from Refs.~\cite{He4Ca40Pb208:Hauser, He4Ca40:Goldberg, He4Ca40:Youngblood}. 
}
 \label{fig:4He+40Ca-scattering}
\end{centering}
\end{figure*}

Figure \ref{fig:4He+208Pb-scattering} shows differential 
cross sections $d \sigma/d \Omega$ as a function of transfer momentum $q$ 
for $^{4}$He scattering from a $^{208}$Pb target in  
$E_{\rm in}/A_{\rm P}=26$--$175$ MeV 
where the experimental data are available. 
The solid and dashed lines stand for the results of the 
Kyushu chiral $g$-matrix DF model with and without 3NF effects, respectively. 
Chiral 3NFs improve the agreement of the theoretical results with 
the experimental data. Particularly for 
$E_{\rm in}/A_{\rm P} \ga 100$ MeV, the agreement is pretty good. 
We can observe the same features also for $^{58}$Ni and $^{40}$Ca targets, 
as shown in Figs. \ref{fig:4He+58Ni-scattering} and 
\ref{fig:4He+40Ca-scattering}, although there is a tendency that the agreement 
becomes better as the target mass increases.

%----------------------
% Figure 4He+208Pb
%----------------------
%----------------------

%----------------------
% Figure 4He+58Ni
%----------------------
%----------------------

%----------------------
% Figure 4He+40Ca
%----------------------
%----------------------

\begin{figure*}[tbp]
\begin{centering}
 \includegraphics[width=0.86\textwidth,clip]{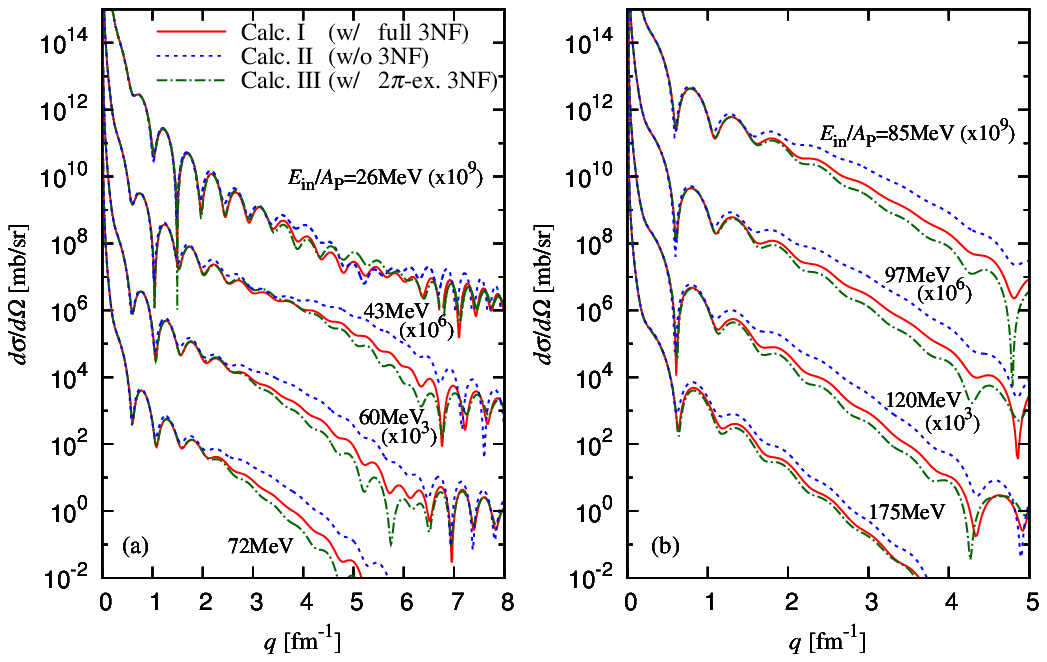}
 \caption{(Color online) 
Effects of Fujita-Miyazawa 2$\pi$-exchange 3NF 
on differential cross sections $d \sigma/d \Omega$ 
for $^{4}$He+$^{58}$Ni scattering, where $q$ is the 
transfer momentum. The solid and dashed lines denote the results of 
calculations I and II, respectively, and 
the dot-dashed line corresponds to the results of calculations III; 
see Sec. \ref{Some basic results of BHF calculations} for 
the definition of $g$-matrix calculations. 
Each cross section is multiplied by the factor shown in the figure.  
}
 \label{fig:4He+58Ni-scattering-diagram-a}
\end{centering}
\end{figure*}

Now we analyze effects of Fujita-Miyazawa 2$\pi$-exchange 3NF 
on differential cross sections $d \sigma/d \Omega$ 
for $^{4}$He+$^{58}$Ni scattering. 
In Fig \ref{fig:4He+58Ni-scattering-diagram-a}, 
the solid, dashed and dot-dashed lines denote the results of 
calculations I, II and III, respectively; 
see Sec. \ref{Some basic results of BHF calculations} for 
the definition of $g$-matrix calculations. 
The difference between calculations I and II 
means effects of all 3NFs, and that between calculations II and III 
corresponds to effects of Fujita-Miyazawa 2$\pi$-exchange 3NF. 
The resultant cross sections show that 
the Fujita-Miyazawa 2$\pi$-exchange 3NF is 
the main contribution of chiral-3NF effects 
on $^{4}$He scattering.

%----------------------
% Figure 4He+58Ni-FM
%----------------------
%----------------------

\begin{figure*}[tbp]
 \begin{centering}
  \includegraphics[width=0.78\textwidth,clip]{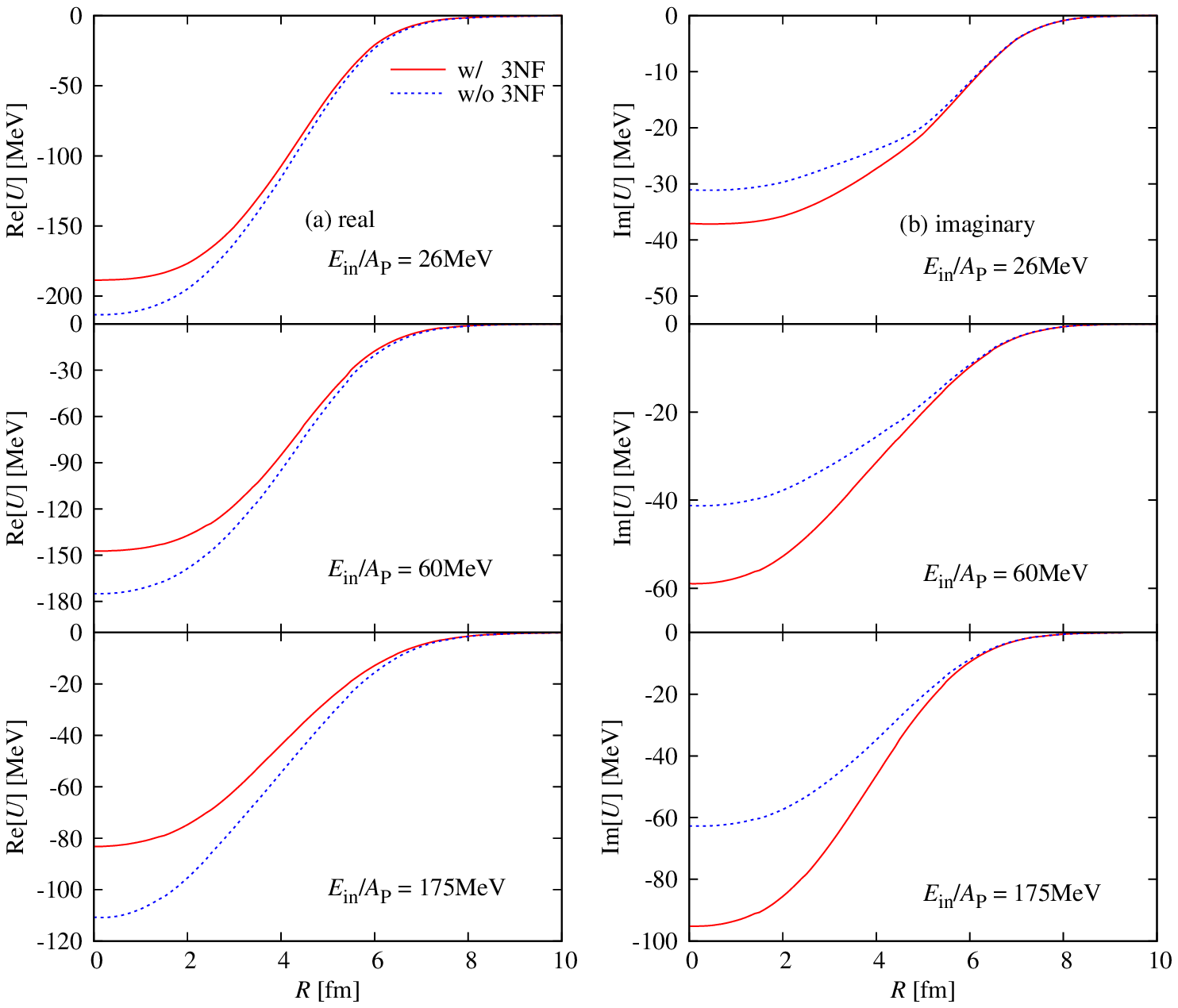}  
 \caption{(Color online) 
Optical potentials $U(R)$ as a function of $R$ 
for $^{4}$He+$^{58}$Ni elastic scattering 
at $E_{\rm in}/A_{\rm P}=$26, 60 and $175$ MeV. 
The solid (dashed) lines denote the optical potentials 
with (without) chiral-3NF effects. 
Panels (a) and (b) represent the real and imaginary 
parts of $U$, respectively. 
}
 \label{fig:He4Ni58-potential}
 \end{centering}
\end{figure*}

Figure \ref{fig:He4Ni58-potential} shows the $R$ dependence of 
the optical potentials $U(R)$ 
for $^{4}$He elastic scattering from a $^{58}$Ni target 
at $E_{\rm in}/A_{\rm P}=$26, 60 and $175$ MeV. 
The solid and dashed lines represent the $U(R)$ 
with and without chiral-3NF effects; 
note that only the central potential is generated by the DF-TDA model. 
As expected, chiral-3NF effects make repulsive and absorptive corrections 
to the optical potentials, and the corrections slightly increase as 
$E_{\rm in}$ goes up; note that the effects hardly depend on $E_{\rm in}$ 
in the peripheral region, $R \approx 6$~fm, that is important for 
the elastic scattering. 
As already mentioned in Sec. \ref{Some basic results of BHF calculations}, 
the repulsive correction mainly comes from 
the Fujita-Miyazawa 2$\pi$-exchange 3NF in its $^{3}$O component, and 
the absorptive correction stems from the $^{3}$E and $^{3}$O components of 
Fujita-Miyazawa 2$\pi$-exchange 3NF.

%----------------------
% Figure optical potential for  He4-scattering 
%----------------------
%----------------------

\begin{figure*}[tbp]
 \begin{centering}
  \includegraphics[width=0.86\textwidth,clip]{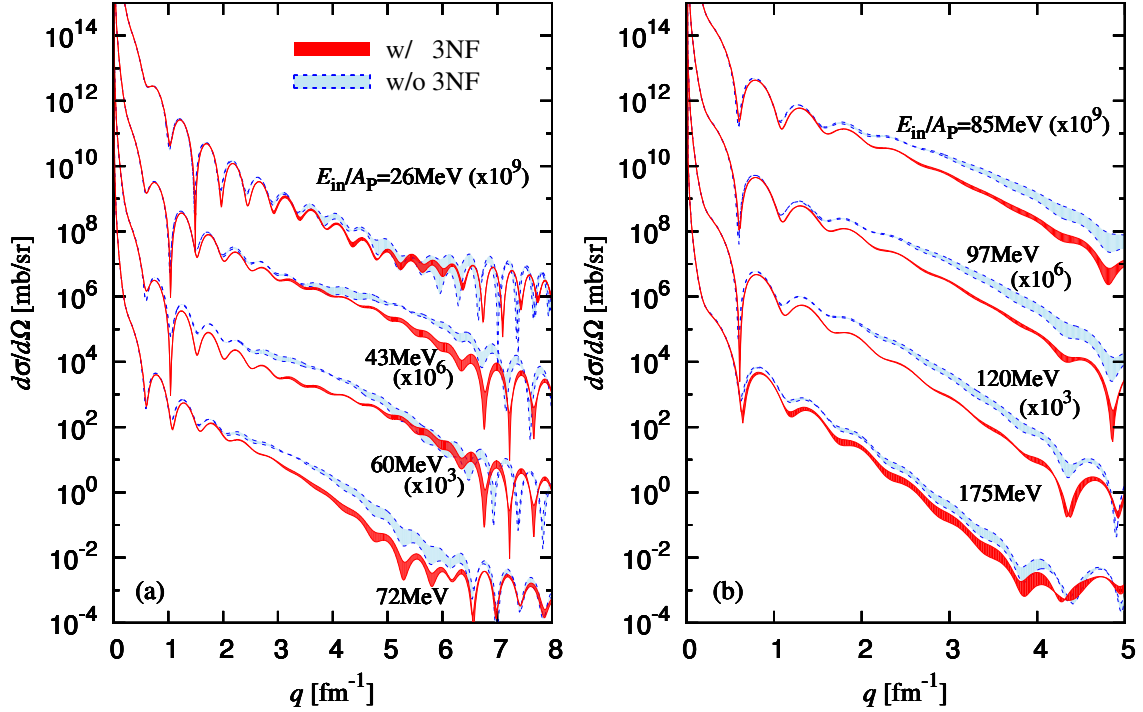}  
 \caption{(Color online) 
  Uncertainty coming from $\Lambda$ dependence of differential cross sections 
  $d \sigma/d \Omega$ for $^{4}$He+$^{58}$Ni elastic scattering. 
  $\Lambda$ dependence is drawn by a hatching for each of 
  2NF and 2NF+3NF calculations, where two cases of $\Lambda=550$ and $450$ MeV 
  are taken. 
  Note that the hatching region surrounded by the solid (dashed) lines 
  corresponds to the uncertainty coming from $\Lambda$ dependence for 
  2NF+3NF (2NF) calculations. 
  }
 \label{fig:He4Pb208-cutoff_prc}
 \end{centering}
\end{figure*}

Figure \ref{fig:He4Pb208-cutoff_prc} shows the uncertainty coming from 
$\Lambda$ dependence of differential cross sections  $d \sigma/d \Omega$ 
for $^{4}$He+$^{58}$Ni elastic scattering. Here 
two cases of $\Lambda=550$ and $450$ MeV are considered.   
$\Lambda$ dependence is shown by a hatching for each of 
2NF and 2NF+3NF calculations; 
note that the hatching region surrounded by solid (dashed) lines 
means the uncertainty coming from $\Lambda$ dependence for 
2NF+3NF (2NF) calculations. 
As expected, $\Lambda$ dependence becomes larger 
as $E_{\rm in}$ increases, 
but the uncertainty coming from $\Lambda$ dependence is still smaller than 
chiral-3NF effects, even at $E_{\rm in}/A_{\rm P}=175$ MeV.

%----------------------
% Figure cutoff He4-scattering 
%----------------------
%----------------------

\begin{figure}[tbp]
\begin{centering}
 \includegraphics[width=0.24\textwidth]{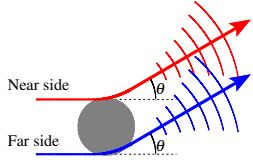}
 \caption{(Color online) 
Illustration of the near/far decomposition. 
}
 \label{Fig:Near/Far-decomposition}
\end{centering}
\end{figure}

The scattering amplitude can be decomposed into the near- 
and far-side components \cite{Fuller:1975}. 
As illustrated in Fig. \ref{Fig:Near/Far-decomposition}, 
these components are well defined, when outgoing waves are generated only 
in the peripheral region of T. 
$^{4}$He scattering on a heavier target is a good case. 
The absorptive correction of chiral-3NF effects 
makes the decomposition more applicable. 
The decomposition is a convenient tool for investigating the interplay 
between differential cross sections $d \sigma/d \Omega$ 
and the real part of $U(R)$. 
The near-side (far-side) outgoing waves are mainly induced by 
repulsive Coulomb (attractive nuclear) force, so that 
very-forward-angle (middle-angle) scattering are dominated by 
the near-side (far-side) components. 
As a consequence of this property, a large interference pattern 
appears in differential cross sections 
at the forward angles where the two components become comparable, 
and the far-side dominance is realized at middle angles 
after the interference pattern. In the middle angle region, 
any repulsive correction to $U(R)$ reduces differential cross sections.

\begin{figure}[tbp]
\begin{centering}
 \includegraphics[width=0.40\textwidth,clip]{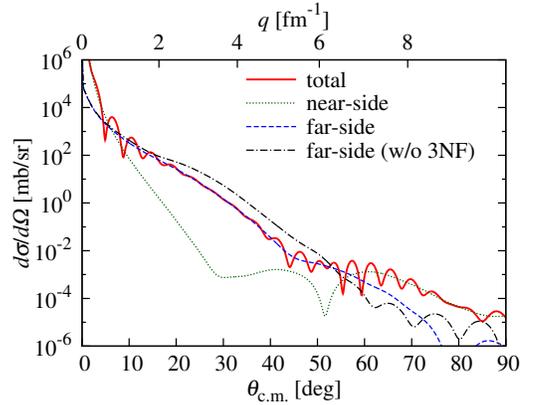}
 \caption{(Color online) 
Near/far decomposition of differential cross sections $d \sigma/d \Omega$ 
for $^{4}$He+$^{58}$Ni scattering at 
$E_{\rm in }/A_{\rm P} = 72$ MeV. 
The dotted (dashed) line stands for the near-side (far-side) cross sections, 
while the solid line denotes differential cross sections before 
the near/far decomposition; here chiral-3NF effects are taken into account. 
The dot-dashed line corresponds to the far-side cross section 
in which chiral 3NFs are switched off. 
}
 \label{Fig:He4Ni58E72-nearfar}
\end{centering}
\end{figure}

Figure \ref{Fig:He4Ni58E72-nearfar} shows the near/far decomposition 
of differential cross sections $d \sigma/d \Omega$ 
for $^{4}$He+$^{58}$Ni scattering at 
$E_{\rm in }/A_{\rm P} = 72$ MeV. 
The dotted and dashed lines represent the near- and far-side cross sections, 
respectively, 
and the solid line denotes differential cross sections before 
the near/far decomposition; here chiral-3NF effects are taken into account. 
The solid line shows a large interference pattern 
at $\theta_{\rm c.m.}= 5$--$15^\circ$, and the solid line agrees with the 
dashed one in $20^\circ \la \theta_{\rm c.m.} \la 40^\circ$. The far-side dominance is thus 
realized in middle angles $20^\circ \la \theta_{\rm c.m.} \la 40^\circ$. 
The far-side dominance in $20^\circ \la \theta_{\rm c.m.} \la 40^\circ$ 
persists, even after chiral 3NFs are switched off. 
The dot-dashed line is the far-side cross section 
in which chiral 3NFs are switched off. 
The repulsive correction coming from chiral 3NFs suppresses 
differential cross sections 
in far-side dominant angles $20^\circ \la \theta_{\rm c.m.} \la 40^\circ$ 
from the dot-dashed line to the solid (dashed) line. 
Thus, chiral-3NF effects become more visible 
in the far-side dominant angle region.

%----------------------
% Figure Near/Far-illust
%----------------------
%----------------------

%----------------------
% Figure He4Ni58E72-nearfar 
%----------------------
%----------------------

\begin{figure}[tbp]
 \begin{center}
  \includegraphics[width=0.39\textwidth,clip]{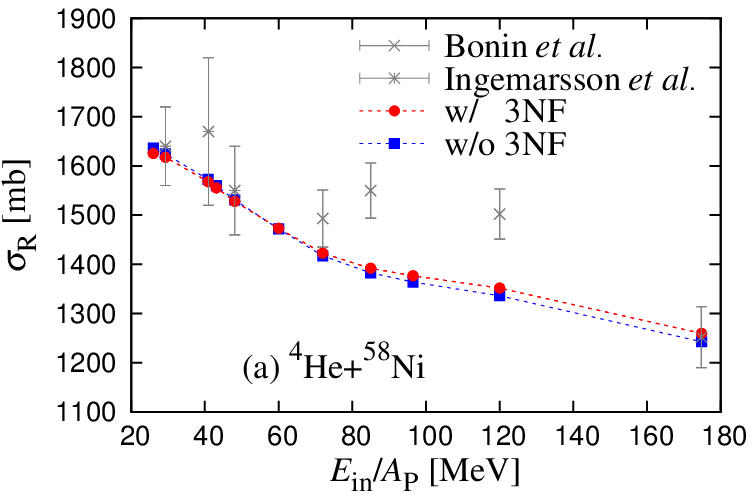}
  \includegraphics[width=0.39\textwidth,clip]{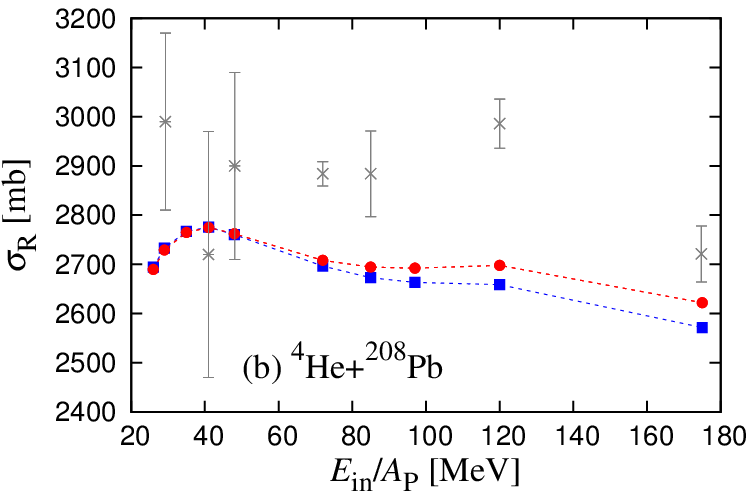}
  \caption{(Color online) 
  Total reaction cross sections $\sigma_{\rm R}$ 
  as a function of $E_{\rm in}$ 
  for $^{4}$He scattering on (a) $^{58}$Ni and (b) $^{58}$Pb targets. 
  Closed circles (squares) denote the results of 
  Kyushu chiral $g$ matrix with (without) 3NF effects. 
  Experimental data are taken from Refs.~\cite{rcs:Bonin, rcs:Ingemarsson}. 
  }
  \label{fig:4He+58Ni-rcs}
 \end{center}
\end{figure}

Finally, we comment on chiral-3NF effects on total reaction cross sections 
$\sigma_{\rm R}$ briefly. 
Radii of stable and unstable nuclei are often determined 
from measured $\sigma_{\rm R}$ with the folding model and/or 
the Glauber model. 
Figure \ref{fig:4He+58Ni-rcs} shows $\sigma_{\rm R}$ 
as a function of $E_{\rm in}/A_{\rm P}$ for $^{4}$He scattering on $^{58}$Ni 
and $^{208}$Pb targets. 
Closed circles (squares) mean the results of 
Kyushu chiral $g$ matrix with (without) 3NF effects. 
The two kinds of results are close to each other, 
indicating that chiral-3NF effects are negligible for $\sigma_{\rm R}$. 
The fact ensures that the determination of nuclear radii from 
measured $\sigma_{\rm R}$ is reliable.

%----------------------
% Figure 4He+58Pb RCS
%----------------------
%----------------------

%Summary
\section{Summary} 
\label{Summary}

We investigated basic properties of chiral 3NFs 
in symmetric nuclear matter 
with positive energies up to 200~MeV by using the BHF method with 
chiral 2NFs of N$^{3}$LO and chiral 3NFs of NNLO in the Bochum-Bonn-J\"{u}lich \cite{Epe05}, 
parameterization, 
and analyzed chiral-3NF effects on $^{4}$He elastic scattering 
from heavier targets $^{208}$Pb, $^{58}$Ni and $^{40}$Ca over 
a wide incident-energy range of 
$30 \la E_{\rm in}/A_{\rm P} \la 200$~MeV 
by the Kyushu chiral $g$-matrix folding model.

First, we summarize the basic properties of chiral 3NFs in symmetric 
nuclear matter with positive energies $E_{\rm in}$ up to 200 MeV: 
\begin{enumerate}
\renewcommand{\labelenumi}{(\arabic{enumi})}
\item 
Chiral 3NFs make the single-particle potential ${\cal U}$ less attractive and 
more absorptive. 
\item 
The repulsive and absorptive corrections slightly increase as $E_{\rm in}$ 
 goes up.
\item
Chiral 3NF effects on ${\cal U}$ mainly come from the 
Fujita-Miyazawa 2$\pi$-exchange 3NF (diagram (a) in Fig. \ref{fig:diagram}). 
More precisely, the repulsion mainly stems from 
the $^{3}$O component of the diagram (a) and the absorption does 
from the $^{3}$O and $^{3}$E components of the diagram (a). 
\end{enumerate}

Properties (1)-(3) persist in the optical potential of $^{4}$He scattering. 
This is natural, since the single-particle potential plays a role of 
the optical potential in nuclear matter. However, it should be noted that 
chiral-3NF effects depend little on $E_{\rm in}$ 
in the peripheral region that is important for the elastic scattering.

Chiral-3NF effects are evident for $^{4}$He scattering in  
$E_{\rm in}/A_{\rm P} \ga 60$ MeV at the middle angles where 
the cross sections are dominated by the far-side component of 
the scattering amplitude. The repulsive correction of chiral 
3NFs reduces  the far-side component and thereby yields 
better agreement with the experimental data. 
Eventually, the Kyushu chiral $g$-matrix DF model reproduces measured 
differential cross sections pretty well, particularly for 
$^{4}$He scattering at $E_{\rm in}/A_{\rm P} \ga 100$ MeV.

All the analyses mentioned above were made with $\Lambda=550$ MeV.
In order to investigate $\Lambda$ dependence in 
nuclear-matter and $^{4}$He-scattering calculations, we take $\Lambda=450$ MeV 
in addition to $\Lambda=550$ MeV. 
The uncertainty coming from $\Lambda$ dependence is smaller than chiral-3NF effects.
There is a tendency that the uncertainty becomes larger as $E_{\rm in}$ 
increases, but it is still smaller than chiral-3NF effects 
even at $E_{\rm in} = 175$~MeV.

Finally, we provide the local version of chiral $g$-matrix 
with a 3-range Gaussian form for the case of $E_{\rm in}=72$ MeV. 
Numerical numbers are presented in Appendix \ref{Appendix}. 
This local version of chiral $g$ matrix strongly encourages us to 
use it for studying various kinds of nuclear reactions.

%Acknowledgement
\section*{Acknowledgements} 

The authors would like to thank K. Ogata and K. Minomo 
for valuable discussions on localization of the chiral $g$-matrix and the reaction analyses. 
This work is supported in part by
by Grant-in-Aid for Scientific Research
(Nos. 25400266, 26400278, 16K05353, and 16J00630)
from Japan Society for the Promotion of Science (JSPS).

\appendix
\section{Parameter set of Kyushu chiral $g$ matrix}
\label{Appendix}
In this Appendix, we provide the central part of 
the Kyushu chiral $g$ matrix, for the case of 
$E_{\rm in}/A_{\rm P}=75$ MeV, 
in a 3-range Gaussian form
\begin{eqnarray}
 g^{ST}(s,k_{\rm F},E_{\rm in}/A_{\rm P})
  =\sum_{i=1}^{3}
  g^{ST}_{i}(k_{\rm F},E_{\rm in}/A_{\rm P})
  e^{-s^2/\lambda_i^2}
\end{eqnarray}
in each $(S, T)$ channel.
The range parameters are fixed to be
$(\lambda_1,\lambda_2,\lambda_3)=(0.4, 0.9, 2.5)$
in units of fm, and the strength parameters 
$\tilde{g}^{ST}_{i}(k_{\rm F},E_{\rm in}/A_{\rm P})$
in units of MeV
, which include chiral 3NF effects,
are tabulated in Tables \ref{singlet-even}--\ref{triplet-odd}
for six cases of the Fermi momentum $k_{\rm F}$. 
We will publish parameter sets of other cases on the website
\cite{Web}. 

%\begin{landscape}
\begin{table*}[h]
 \begin{center}
  \begin{minipage}{\textwidth}
   \caption{
   Singlet-even ($S=0,T=1$) component of Kyushu chiral $g$ matrix
   for the incident energy $E_{\rm in}/A_{\rm P}=75$ MeV. 
   The range parameters are fixed to be 
   $(\lambda_1,\lambda_2,\lambda_3)=(0.4, 0.9, 2.5)$ in units of fm. 
   Entries are in MeV, but $k_{\rm F}$ is presented in units of fm$^{-1}$. 
   }
   \begin{tabular}{ccccccccc}
    \hline
    \hline
    &&\multicolumn{3}{c}{real part}
    &&\multicolumn{3}{c}{imaginary part}\\
    \cline{3-5}\cline{7-9}
    $k_{\rm F}$
    &&$i=1$&$i=2$&$i=3$
    &&$i=1$&$i=2$&$i=3$
    \\
    \cline{1-1}\cline{3-3}\cline{4-4}\cline{5-5}
    \cline{7-7}\cline{8-8}\cline{9-9}
    $0.00$&&    1.78627$\times 10^{3}$&   -2.70833$\times 10^{2}$&   -4.08777$\times 10^{0}$&&    2.16654$\times 10^{3}$&   -3.13038$\times 10^{2}$&    1.39295$\times 10^{0}$\\
    $0.60$&&    1.47941$\times 10^{3}$&   -2.47638$\times 10^{2}$&   -3.68242$\times 10^{0}$&&    1.13963$\times 10^{3}$&   -1.71777$\times 10^{2}$&    8.14716$\times 10^{-1}$\\
    $0.80$&&    1.36782$\times 10^{3}$&   -2.39203$\times 10^{2}$&   -3.53502$\times 10^{0}$&&    7.66206$\times 10^{2}$&   -1.20410$\times 10^{2}$&    6.04451$\times 10^{-1}$\\
    $1.10$&&    1.20044$\times 10^{3}$&   -2.26551$\times 10^{2}$&   -3.31392$\times 10^{0}$&&    2.06075$\times 10^{2}$&   -4.33587$\times 10^{1}$&    2.89052$\times 10^{-1}$\\
    $1.20$&&    1.09716$\times 10^{3}$&   -2.13700$\times 10^{2}$&   -3.19454$\times 10^{0}$&&    1.25689$\times 10^{2}$&   -2.88651$\times 10^{1}$&    1.74997$\times 10^{-1}$\\
    $1.30$&&    9.54436$\times 10^{2}$&   -1.94839$\times 10^{2}$&   -3.14638$\times 10^{0}$&&    6.90248$\times 10^{1}$&   -1.93121$\times 10^{1}$&    9.44602$\times 10^{-2}$\\
    $1.40$&&    7.65583$\times 10^{2}$&   -1.68406$\times 10^{2}$&   -3.19684$\times 10^{0}$&&    1.85011$\times 10^{1}$&   -1.05755$\times 10^{1}$&    2.29847$\times 10^{-2}$\\
    $1.50$&&    5.88265$\times 10^{2}$&   -1.44596$\times 10^{2}$&   -3.21499$\times 10^{0}$&&   -1.57455$\times 10^{0}$&   -7.13793$\times 10^{0}$&   -1.93808$\times 10^{-3}$\\
   \end{tabular}
   \label{singlet-even}
  \end{minipage}
  
  \begin{minipage}{\textwidth}
   \caption{
   Triplet-even ($S=1,T=0$) component of Kyushu chiral $g$ matrix
   for the incident energy $E_{\rm in}/A_{\rm P}=75$ MeV. 
   See Table \ref{singlet-even} for the detail.
   }
   \begin{tabular}{ccccccccc}
    \hline
    \hline
    &&\multicolumn{3}{c}{real part}
    &&\multicolumn{3}{c}{imaginary part}\\
    \cline{3-5}\cline{7-9}
    $k_{\rm F}$
    &&$i=1$&$i=2$&$i=3$
    &&$i=1$&$i=2$&$i=3$
    \\
    \cline{1-1}\cline{3-3}\cline{4-4}\cline{5-5}
    \cline{7-7}\cline{8-8}\cline{9-9}
    $0.00$&&    1.25135$\times 10^{3}$&   -2.14233$\times 10^{2}$&   -4.23044$\times 10^{0}$&&    2.83713$\times 10^{3}$&   -4.39936$\times 10^{2}$&   -7.11017$\times 10^{-1}$\\
    $0.60$&&    1.26094$\times 10^{3}$&   -2.62345$\times 10^{2}$&   -3.04629$\times 10^{0}$&&    1.96817$\times 10^{3}$&   -3.18594$\times 10^{2}$&   -4.36898$\times 10^{-1}$\\
    $0.80$&&    1.26443$\times 10^{3}$&   -2.79841$\times 10^{2}$&   -2.61569$\times 10^{0}$&&    1.65218$\times 10^{3}$&   -2.74470$\times 10^{2}$&   -3.37218$\times 10^{-1}$\\
    $1.10$&&    1.26966$\times 10^{3}$&   -3.06084$\times 10^{2}$&   -1.96978$\times 10^{0}$&&    1.17821$\times 10^{3}$&   -2.08284$\times 10^{2}$&   -1.87698$\times 10^{-1}$\\
    $1.20$&&    1.13634$\times 10^{3}$&   -2.90035$\times 10^{2}$&   -1.65944$\times 10^{0}$&&    9.55010$\times 10^{2}$&   -1.67104$\times 10^{2}$&   -3.11895$\times 10^{-1}$\\
    $1.30$&&    9.50006$\times 10^{2}$&   -2.63272$\times 10^{2}$&   -1.57370$\times 10^{0}$&&    8.52716$\times 10^{2}$&   -1.48093$\times 10^{2}$&   -4.61621$\times 10^{-1}$\\
    $1.40$&&    5.98320$\times 10^{2}$&   -2.11212$\times 10^{2}$&   -1.85361$\times 10^{0}$&&    8.38254$\times 10^{2}$&   -1.45648$\times 10^{2}$&   -5.53945$\times 10^{-1}$\\
    $1.50$&&    4.87230$\times 10^{2}$&   -1.95265$\times 10^{2}$&   -1.99886$\times 10^{0}$&&    8.31225$\times 10^{2}$&   -1.43294$\times 10^{2}$&   -6.62903$\times 10^{-1}$\\
   \end{tabular}
   \label{triplet-even}
  \end{minipage}
  \begin{minipage}{\textwidth}
   \caption{ 
   Singlet-odd ($S=0,T=0$) component of Kyushu chiral $g$ matrix
   for the incident energy $E_{\rm in}/A_{\rm P}=75$ MeV. 
   See Table \ref{singlet-even} for the detail. 
   }
   \begin{tabular}{ccccccccc}
    \hline
    \hline
    &&\multicolumn{3}{c}{real part}
    &&\multicolumn{3}{c}{imaginary part}\\
   \cline{3-5}\cline{7-9}
    $k_{\rm F}$
    &&$i=1$&$i=2$&$i=3$
    &&$i=1$&$i=2$&$i=3$
    \\
    \cline{1-1}\cline{3-3}\cline{4-4}\cline{5-5}
    \cline{7-7}\cline{8-8}\cline{9-9}
    $0.00$&&    1.17797$\times 10^{3}$&    1.54048$\times 10^{1}$&    9.23703$\times 10^{0}$&&    6.01921$\times 10^{2}$&   -6.69649$\times 10^{1}$&   -3.21021$\times 10^{-1}$\\
    $0.60$&&    2.92329$\times 10^{2}$&    9.73676$\times 10^{1}$&    8.54387$\times 10^{0}$&&    4.56013$\times 10^{2}$&   -5.19773$\times 10^{1}$&   -1.83729$\times 10^{-1}$\\
    $0.80$&&   -2.97222$\times 10^{1}$&    1.27172$\times 10^{2}$&    8.29181$\times 10^{0}$&&    4.02955$\times 10^{2}$&   -4.65273$\times 10^{1}$&   -1.33805$\times 10^{-1}$\\
    $1.10$&&   -5.12800$\times 10^{2}$&    1.71879$\times 10^{2}$&    7.91372$\times 10^{0}$&&    3.23369$\times 10^{2}$&   -3.83522$\times 10^{1}$&   -5.89192$\times 10^{-2}$\\
    $1.20$&&   -6.98327$\times 10^{2}$&    1.94038$\times 10^{2}$&    7.80370$\times 10^{0}$&&    3.02061$\times 10^{2}$&   -3.67462$\times 10^{1}$&   -4.49149$\times 10^{-2}$\\
    $1.30$&&   -8.68590$\times 10^{2}$&    2.13672$\times 10^{2}$&    7.65323$\times 10^{0}$&&    3.08332$\times 10^{2}$&   -3.85626$\times 10^{1}$&   -3.61166$\times 10^{-2}$\\
    $1.40$&&   -1.21630$\times 10^{3}$&    2.43679$\times 10^{2}$&    7.28544$\times 10^{0}$&&    3.39578$\times 10^{2}$&   -4.29750$\times 10^{1}$&   -1.70789$\times 10^{-2}$\\
    $1.50$&&   -1.35278$\times 10^{3}$&    2.58695$\times 10^{2}$&    7.07301$\times 10^{0}$&&    3.38596$\times 10^{2}$&   -4.29790$\times 10^{1}$&   -8.86515$\times 10^{-3}$\\
   \end{tabular}
   \label{singlet-odd}
  \end{minipage}
  
  \begin{minipage}{\textwidth}
   \caption{
   Triplet-odd ($S=1,T=1$) component of Kyushu chiral $g$ matrix
   for the incident energy $E_{\rm in}/A_{\rm P}=75$ MeV. 
   See Table \ref{singlet-even} for the detail.
   }
   \begin{tabular}{ccccccccc}
    \hline
    \hline
    &&\multicolumn{3}{c}{real part}
    &&\multicolumn{3}{c}{imaginary part}\\
    \cline{3-5}\cline{7-9}
    $k_{\rm F}$
    &&$i=1$&$i=2$&$i=3$
    &&$i=1$&$i=2$&$i=3$
    \\
    \cline{1-1}\cline{3-3}\cline{4-4}\cline{5-5}
    \cline{7-7}\cline{8-8}\cline{9-9}
    $0.00$&&    1.48087$\times 10^{3}$&   -1.17015$\times 10^{2}$&    4.09818$\times 10^{-1}$&&    6.26964$\times 10^{2}$&   -5.48253$\times 10^{1}$&   -2.34394$\times 10^{-1}$\\
    $0.60$&&    9.89803$\times 10^{2}$&   -7.76051$\times 10^{1}$&    3.97949$\times 10^{-1}$&&    4.90110$\times 10^{2}$&   -4.36088$\times 10^{1}$&   -1.27683$\times 10^{-1}$\\
    $0.80$&&    8.11232$\times 10^{2}$&   -6.32740$\times 10^{1}$&    3.93633$\times 10^{-1}$&&    4.40345$\times 10^{2}$&   -3.95300$\times 10^{1}$&   -8.88792$\times 10^{-2}$\\
    $1.10$&&    5.43375$\times 10^{2}$&   -4.17774$\times 10^{1}$&    3.87159$\times 10^{-1}$&&    3.65697$\times 10^{2}$&   -3.34119$\times 10^{1}$&   -3.06733$\times 10^{-2}$\\
    $1.20$&&    3.30284$\times 10^{2}$&   -2.25888$\times 10^{1}$&    3.93880$\times 10^{-1}$&&    3.57729$\times 10^{2}$&   -3.29841$\times 10^{1}$&   -1.48011$\times 10^{-2}$\\
    $1.30$&&    1.31873$\times 10^{2}$&   -4.61127$\times 10^{0}$&    3.96357$\times 10^{-1}$&&    3.94174$\times 10^{2}$&   -3.67176$\times 10^{1}$&   -2.33502$\times 10^{-3}$\\
    $1.40$&&   -9.73856$\times 10^{1}$&    1.42601$\times 10^{1}$&    3.20402$\times 10^{-1}$&&    4.67099$\times 10^{2}$&   -4.37829$\times 10^{1}$&    2.27706$\times 10^{-2}$\\
    $1.50$&&   -2.46017$\times 10^{2}$&    2.76700$\times 10^{1}$&    3.11786$\times 10^{-1}$&&    5.04428$\times 10^{2}$&   -4.77850$\times 10^{1}$&    3.25825$\times 10^{-2}$\\
   \end{tabular}
   \label{triplet-odd}
  \end{minipage}
 \end{center}
\end{table*}
%\end{landscape}

%%--------------------------------------------------------------------%%
%%                           References                               %%
%%--------------------------------------------------------------------%%


\begin{thebibliography}{00}



%%% Spin-Orbit Coupling in Heavy Nuclei
\bibitem{Fuj57}
J.~Fujita and H.~Miyazawa,
Prog. Theor. Phys.~{\bf 17}, 360 (1957); 
{\it ibid}.~{\bf 17}, 366 (1957).


%%% phenomenological 3NF
\bibitem{Wiringa2002}
R.~B.~Wiringa and S.~C.~Pieper, 
Phys. Rev. Lett. {\bf 89}, 182501 (2002). 

\bibitem{Wiringa1988}
R.~B.~Wiringa, V.~Fiks, and A.~Fabrocini, 
Phys. Rev. C {\bf 38}, 1010 (1988).


%%% chiral EFT -- review ---
\bibitem{Epelbaum-review-2009}
E.~Epelbaum, H.-W.~Hammer, and Ulf-G.~Mei{\ss}ner, 
Rev. Mod. Phys.~{\bf 81}, 1773 (2009).
\bibitem{Machleidt-2011}
R.~Machleidt and D.~R.~Entem, Phys. Rep.~{\bf 503}, 1 (2011).

%%% roles of chiral 3NF
% nuclear matter and light nuclei
\bibitem{Hammer13}
H.~-W.~Hammer, A.~Nogga, and A.~Schwenk, 
Rev. Mod. Phys.~{\bf 85}, 197 (2013).
% light nuclei
\bibitem{Kalantar-2012}
\mbox{N.~Kalantar-Nayestanaki, E.~Epelbaum}, 
\mbox{J.~G.~Messchendorp, and A.~Nogga, 
Rep. Prog.} Phys.~{\bf 75}, 016301 (2012).
% ab initio
\bibitem{Holt14}
J.~D.~Holt, J.~Men\'{e}ndez, J.~Simonis, and A.~Schwenk,
Phys. Rev. C~{\bf 90}, 024312 (2014).
\bibitem{Ekstrom:2015rta} 
A.~Ekstr\"{o}m {\it et al}., 
Phys.\ Rev.\ C~{\bf 91}, 051301(R) (2015).
% nuclear matter
\bibitem{HEB11} 
K.~Hebeler, S.~K.~Bogner, R.~J.~Furnstahl, A.~Nogga, and A.~Schwenk,
Phys. Rev. C~{\bf 83}, 031301(R) (2011).

\bibitem{Samm12} 
F.~Sammarruca, B.~Chen, L.~Coraggio, N.~Itaco, and R.~Machleidt,
Phys. Rev. C~{\bf  86}, 054317 (2012). 
\bibitem{Koh13}
M.~Kohno, 
Phys. Rev. C~{\bf 88}, 064005 (2013).
\bibitem{Koh15}
M.~Kohno,
Prog. Theor. Exp. Phys.~{\bf 123}D02 (2015).
\bibitem{Kru13}
T.~Kr\"{u}ger, I.~Tews, K.~Hebeler, and A.~Schwenk,
Phys. Rev. C~{\bf 88}, 025802 (2013).
\bibitem{Dri14}
C.~Drischler, V.~Som\`{a}, and A. Schwenk,
Phys. Rev. C~{\bf 89}, 025806 (2014).
\bibitem{Kru15}
T.~Kr\"{u}ger, K.~Hebeler, A.~Schwenk,
Phys. Lett. B~{\bf 744}, 18 (2015).

%%% chiral 4NF effects in nuclear matter
\bibitem{Kai12}
N.~Kaiser, Eur. Phys. J. A~{\bf 48}, 135 (2012).

\bibitem{Kaiser:2015lsa} 
  N.~Kaiser and R.~Milkus,
  %``Reducible chiral four-body interactions in nuclear matter,''
  Eur.\ Phys.\ J.\ A {\bf 52}, no. 1, 4 (2016). 

\bibitem{Sekig14}
K.~Sekiguchi {\it et al}., Phys. Rev. C~{\bf 89}, 064007 (2014). 


%%% g-matrix folding model
\bibitem{Brieva-Rook}
F.~A.~Brieva and J.~R.~Rook, Nucl. Phys. A~{\bf 291}, 299 (1977);
{\it ibid.}~291, 317 (1977); {\it ibid.}~297, 206 (1978).

%%% g-matrix folding model
\bibitem{Amos}
K.~Amos, P.~J.~Dortmans, H.~V.~von Geramb, 
S.~Karataglidis, and J.~Raynal, 
in \textit{Advances in Nuclear Physics}, edited by
J.~W.~Negele and E.~Vogt(Plenum, New York, 2000) Vol.~25, p. 275.

\bibitem{CEG07}
T.~Furumoto, Y.~Sakuragi, and Y.~Yamamoto, 
Phys. Rev. C~{\bf 78}, 044610 (2008).
\bibitem{Raf13}
S.~Rafi, M.~Sharma, D.~Pachouri, W.~Haider, and Y.~K.~Gambhir,
Phys. Rev. C~{\bf 87}, 014003 (2013).
\bibitem{MP}
Y.~Yamamoto, T.~Furumoto, N.~Yasutake, and \mbox{Th.~A.~Rijken}, 
Phys. Rev. C~{\bf 88}, 022801 (2013).

%%% chiral g-matrix folding model
\bibitem{Toyokawa:2015zxa} 
M.~Toyokawa, M.~Yahiro, T.~Matsumoto, K.~Minomo, K.~Ogata and M.~Kohno,
Phys.\ Rev.\ C~{\bf 92}, 024618 (2015).

%%% 3NF eff. for AA scattering
\bibitem{CEG07-2}
T.~Furumoto, Y.~Sakuragi, and Y.~Yamamoto, 
Phys. Rev. C~{\bf 80}, 044614 (2009).



%%% Chiral 3NF effects on elastic scattering
\bibitem{Toyokawa:2014yma} 
M.~Toyokawa, K.~Minomo, M.~Kohno and M.~Yahiro,
J. Phys. G~{\bf 42}, 025104 (2015).
\bibitem{Minomo:2014eqa} 
K.~Minomo, M.~Toyokawa, M.~Kohno and M.~Yahiro, 
Phys.\ Rev.\ C {\bf 90}, 051601(R) (2014).


%%% Bonn-B
% \bibitem{BonnB}
% R.~Machleidt, K.~Holinde, and Ch.~Elster,
% Phys. Rep.~{\bf 149}, 1 (1987).

%%% 4He scattering
\bibitem{Egashira:2014zda} 
K.~Egashira, K.~Minomo, M.~Toyokawa, T.~Matsumoto and M.~Yahiro, 
Phys. Rev. C~{\bf 89}, 064611 (2014). 

%%% Erratum:
\bibitem{Koh:Err}
M.~Kohno, Phys. Rev. C {\bf 96}, 059903(E) (2017). 


%%% 3He scattering
\bibitem{Toyokawa:2015fva} 
M.~Toyokawa, T.~Matsumoto, K.~Minomo and M.~Yahiro, 
Phys.\ Rev.\ C~{\bf 91}, 064610 (2015).


%%% Second-order perturbation for nuclear matter
\bibitem{Hol13}
J.~W.~Holt, N.~Kaiser, G.~A.~Miller, and W.~Weise,
Phys. Rev. C~{\bf 88}, 024614 (2013).

%%% low-energy constants
\bibitem{Epe05}
E.~Epelbaum, W.~Gl\"{o}ckle, and Ulf-G.~Mei\ss ner,
Nucl. Phys. A~{\bf 747}, 362 (2005).


% \bibitem{Entem-2003}
% D.~R.~Entem and R.~Machleidt, Phys. Rev. C={\bf 68}, 041001 (2003).

%%% relation between cD and cE
% \bibitem{Nogga-2006}
% A.~Nogga, P.~Navr\'{a}til, B.~R.~Barrett, and J.~P.~Vary,
% Phys. Rev. C~{\bf 73}, 064002 (2006). 

%%% attractive effect of chiral 3NFs for light nulcei
% \bibitem{Maris-2013}
% P.~Maris, J.~P.~Vary, and P.~Navr\'{a}til, 
% Phys. Rev. C~{\bf 87}, 014327 (2013).

%%% localization of g-matrices
\bibitem{von-Geramb-1991} 
H.~V.~von Geramb, K.~Amos, L.~Berge, S.~Br\"autigam, H.~Kohlhoff and 
A.~Ingemarsson, 
Phys. Rev. C~{\bf 44}, 73 (1991). 
\bibitem{Amos-1994}
P.~J.~Dortmans and K.~Amos, 
Phys. Rev. C~{\bf 49}, 1309 (1994)

%%% n+p at 150 MeV
\bibitem{np:Measday}
D.~F.~Measday, 
Phys. Rev.~{\bf 142}, 584 (1966).

%%% 
\bibitem{Izu80}
T. Izumoto, S. Krewald, and A. Faessler,
Nucl. Phys. A~{\bf 341}, 319 (1980).
\bibitem{Yahiro-Glauber}
M.~Yahiro, K.~Minomo, K.~Ogata, and M.~Kawai,
Prog. Theor. Phys.~{\bf 120}, 767 (2008).


\bibitem{Tang1978}
Y.~C.~Tang, M.~LeMere, and D.~R.~Thompson, 
Phys. Rep. {\bf 47}, 167 (1978). 

\bibitem{Aoki1982}
K.~Aoki and H.~Horiuchi, 
Prog. Theor. Phys.~{\bf 69}, 857 (1983), and references therein. 


%%% accuracy of localization
\bibitem{Hag06}
K. Hagino, T. Takehi, and N. Takigawa,
Phys. Rev. C~{\bf 74}, 037601 (2006).
\bibitem{Minomo:2009ds}
K.~Minomo, K.~Ogata, M.~Kohno, Y.~R.~Shimizu, and M.~Yahiro,
J. Phys. G~{\bf 37}, 085011 (2010).


%%DFmodel
\bibitem{Khoa94}
D.~T.~Khoa, W.~von Oertzen, and H.~G.~Bohlen, 
Phys. Rev. C {\bf 49}, 1652 (1994).

\bibitem{Khoa97}
D.~T.~Khoa, G.~R.~Satchler, and W.~von Oertzen, 
Phys. Rev. C{\bf 56}, 954 (1997). 
  
%%% Gogny force
\bibitem{GognyD1S}
J.~F.~Berger, M.~Girod, and D.~Gogny,
Comput. Phys. Commun.~{\bf 63}, 365 (1991). 

%%% removal of c.m. motion
\bibitem{Sum12}
T.~Sumi {\it et al}., 
Phys. Rev. C~{\bf 85}, 064613 (2012). 

%%% charge densities from electron scattering
\bibitem{phen-density}
H.~de Vries, C.~W.~de Jager, and C.~de Vries,
At. Data Nucl. Data Tables~{\bf 36}, 495 (1987).

%%% unfolding of charge density
\bibitem{Singhal}
\mbox{R.~P.~Singhal, M.~W.~S.~Macauley, and} 
\mbox{P.~K.~A.~De~Witt~Huberts, 
Nucl. Instrum. and} Method~{\bf 148}, 113 (1978).

%%% experimental data
%%4He+208Pb
%dcs:E=26
\bibitem{He4Ca40Pb208:Hauser}
G.~Hauser, R.~L\"ohken, H.~Rebel, G.~Schatz, G.~W.~Schweimer, 
and J.~Specht, 
Nucl. Phys. A~{\bf 128}, 81 (1969). 
%dcs:E=35
\bibitem{He4Pb208:Goldberg}
D.~A.~Goldberg, S.~M.~Smith, H.~G.~Pugh, P.~G.~Roos, 
and N.~S.~Wall, 
Phys. Rev. C~{\bf 7}, 1938 (1973). 
%dcs:E=72,85,120,175
\bibitem{He4Ni58Pb208:Bonin}
B.~Bonin {\it et al}., 
Nucl. Phys. A~{\bf 445}, 381 (1985).
%dcs:E=97
\bibitem{He4Pb208:Uchida}
M.~Uchida {\it et al}., 
Phys. Rev. C~{\bf 69}, 051301 (2004). 

%%4He+58Ni
%dcs:E=26
\bibitem{He4Ni58:Rebel}
H.~Rebel, R.~L\"ohken, G.~W.~Schweimer, G.~Schatz, 
and G.~Hauser, 
Z. Phys~{\bf 256}, 258 (1972).
%dcs:E=43
\bibitem{He4Ni58:Albinski}
J.~Albi\'nski {\it et al}., 
Nucl. Phys. A~{\bf 445}, 477 (1985). 
%dcs:E=60
\bibitem{He4Ni58:Clark}
H.~L.~Clark, Y.-W.~Lui, and D.~H.~Youngblood, 
Nucl. Phys. A~{\bf 589}, 416 (1995).
\bibitem{He4Ni58:Lui}
Y.-W.~Lui, D.~H.~Youngblood, H.~L.~Clark, Y.~Tokimoto, 
and B.~John, 
Phys. Rev. C~{\bf 73}, 014314 (2006).
%dcs:E=72,85,120,175
%--> He4Ni58Pb208:Bonin
%dcs:E=97
\bibitem{He4Ni58:Nayak}
B.~K.~Nayak {\it et al}., 
Phys. Lett. B~{\bf 637}, 43 (2006). 

%%4He+40Ca
%dcs:E=26
%--> He4Ca40Pb208:Hauser
%dcs:E=35
\bibitem{He4Ca40:Goldberg}
D.~A.~Goldberg, S.~M.~Smith, and G.~F.~Burdzik, 
Phys. Rev. C~{\bf 10}, 1362 (1974). 
%dcs:E=60
\bibitem{He4Ca40:Youngblood}
D.~H.~Youngblood, Y.~-W.~Lui, and H.~L.~Clark, 
Phys. Rev. C {\bf 55}, 2811 (1997). 

%%% Near/Far decomposition
\bibitem{Fuller:1975}
R.~C.~Fuller, Phys. Rev. C~{\bf 12}, 1561 (1975). 


%%4He+58Ni
%%rcs
\bibitem{rcs:Bonin}
B.~Bonin {\it et al}., 
Nucl. Phys. A~{\bf 445}, 381(1985). 
\bibitem{rcs:Ingemarsson}
A.~Ingemarsson {\it et al}., 
Nucl. Phys. A~{\bf 676}, 3 (2000). 


%%%Kakuri Web
\bibitem{Web}
http://www.nt.phys.kyushu-u.ac.jp






\end{thebibliography}
\end{document}